\documentstyle[preprint,aps,psfig,epsfig]{revtex}
\tighten

\begin{document}
\def\eqn#1{Eq.$\,$#1}
\def\mb#1{\setbox0=\hbox{$#1$}\kern-.025em\copy0\kern-\wd0
\kern-0.05em\copy0\kern-\wd0\kern-.025em\raise.0233em\box0}
\draft
\preprint{}
\title{Kinetic theory of point vortices: diffusion coefficient and systematic
drift}
\author{P.H. CHAVANIS }
\address{Laboratoire de Physique Quantique,
Universit\'e Paul Sabatier, 118 route de Narbonne 31062 Toulouse, France}
\address{Institute for Theoretical Physics, University of California, Santa
Barbara, California}
\date{Phys. Rev. E, to appear August 2001}
\maketitle

\begin{abstract}

\end{abstract}

We develop a kinetic theory for point vortices in two-dimensional
hydrodynamics.  Using standard projection operator technics, we derive
a Fokker-Planck equation describing the relaxation of a ``test''
vortex in a bath of ``field'' vortices at statistical equilibrium. The
relaxation is due to the combined effect of a diffusion and a
drift. The drift is shown to be responsible for the organization of
point vortices at negative temperatures. A description that goes
beyond the thermal bath approximation is attempted. A new kinetic
equation is obtained which respects all conservation laws of the point
vortex system and satisfies a H-theorem. Close to equilibrium this
equation reduces to the ordinary Fokker-Planck equation.

\section{Introduction}
\label{sec_introduction}

It is often useful in two-dimensional turbulence to approximate a continuous
vorticity field by a cloud of point vortices. The main interest is that such a
system is Hamiltonian \cite{kirchhoff} and can be studied by rather ordinary
statistical mechanics. This was first considered by Onsager \cite{onsager} who
showed qualitatively the existence of negative temperature states at which
vortices cluster. He could therefore explain the occurence of large scales
vortices (or ``supervortices'') often observed in nature. This was a remarkable
anticipation since observations were very scarce at that time. His work was
pursued by Joyce \& Montgomery \cite{jm} and  Lundgren \& Pointin \cite{lp} who
introduced  a mean-field approximation and obtained explicit results for the
equilibrium state. They derived in particular a Maxwell-Boltzmann statistics for
the distribution of point vortices at equilibrium. 

Less is known concerning the relaxation towards equilibium. In fact, the
evolution of the $N$-particle distribution function is governed by a Liouville
equation but this equation contains too much information to be of practical use.
One is more interested by the evolution of the one-particle distribution
function $P({\bf r},t)$ which gives the probability that a point vortex be found
in ${\bf r}$ at time $t$. In Ref. \cite{chav98}, we have described the
relaxation of $P({\bf r},t)$ towards the Boltzmann distribution in terms of a
phenomenological Fokker-Planck equation. In this approach, the vortices have a
diffusive motion due to random fluctuations and they experience in addition a
{\it systematic drift} (Chavanis, 1998) directed along the background density
gradient. Physically, the drift is the result of a polarization process and its
mathematical expression  can be determined with a linear response theory
\cite{chav98}. It is found that the drift is ``attractive'' at negative
temperatures so the vortices cluster into macrovortices in agreement with
Onsager's thermodynamical approach. At equilibrium, the drift balances the
scattering and maintains a non trivial vortex distribution (the Boltzmann
distribution) providing a dynamical explanation for the persistence of
clustering. 

In this paper, we justify our phenomenological model by deriving the
Fokker-Planck equation directly from the Liouville equation, using projection
operator technics \cite{wp}. These methods are standard in statistical mechanics
but they are applied here for the first time to a system of vortices. We first
consider the relaxation of a ``test'' vortex in a thermal bath in which the
``field'' vortices are in statistical equilibrium. In this approximation, the
Fokker-Planck equation appears in its usual form with a diffusion term and a
drift term. The drift coefficient is connected to the diffusion coefficient and
to the temperature of the bath $1/\beta$ by an Einstein formula. The diffusion
coefficient is expressed as a Kubo formula, i.e. as the integral of the velocity
autocorrelation function. Using an approximation in which the vortices are
advected by the equilibrium flow, we find that the
autocorrelation function decays like $t^{-2}$ for large times \cite{chav98}.
This is a slow decay but it ensures the convergence of the diffusion
coefficient. We also derive non Markovian equations that keep track of memory
effects. Then, we relax the thermal bath approximation and derive a generalized
kinetic equation for our vortex system. This integrodifferential equation
satisfies all conservation laws of the point vortex system and increases the
Boltzmann entropy (H-theorem). The relaxation towards equilibrium is due to a
condition of resonance. If this condition is not satisfied, the system can
remain frozen in a sort of ``metastable'' equilibrium. By contrast, if the
system is sufficiently resonant, it will converge towards the maximum entropy
state described by the Boltzmann distribution. Close to equilibrium, our
generalized kinetic equation reduces to the ordinary Fokker-Planck equation. 

The methods developed in this paper are inspired by those introduced in plasma
physics and stellar dynamics \cite{kandrup}. In particular, the {\it systematic
drift} \cite{chav98} of a point vortex is the counterpart of the {\it dynamical
friction} \cite{chandrasekhar} experienced by a star in a stellar system.
Further analogies between 2D vortices and stellar systems are discussed in the
paper and in  \cite{csr,chavthese,chavNY,hayama,cs,cs2}. Other kinetic theories of point vortices have been developed in Refs. \cite{nazarenko,marmanis} in a different context. A good review on point vortex dynamics is given by Newton \cite{newton}.

\section{Statistical mechanics of point vortices}
\label{sec_statmech}

\subsection{The point vortex model}
\label{sec_model}

In a two-dimensional incompressible fluid, the velocity field ${\bf u}$ is
divergenceless and can be written in terms of a stream function $\psi$ in the
form  
\begin{equation}
{\bf u}=-{\bf z}\wedge \nabla\psi,
\label{div}
\end{equation} 
where ${\bf z}$ is a unit vector normal to the flow. The stream function is
related to the vorticity $\omega {\bf z}=\nabla\wedge {\bf u}$ by the Poisson
equation  
\begin{equation}
\omega=-\Delta\psi
\label{poisson}
\end{equation}  
obtained by taking the curl of equation (\ref{div}). The impermeability
condition implies that $\psi$ is constant on the boundary and we shall take
$\psi=0$ by convention.

We shall consider the situation in which the velocity is created by a collection
of $N$ point vortices of equal circulation $\gamma$. In that case, the vorticity
field can be expressed as a sum of $\delta$-functions in the form
\begin{equation}
\omega ({\bf r},t)=\sum_{i=1}^{N}\gamma\delta ({\bf r}-{\bf r}_{i}(t)),
\label{omegadelta}
\end{equation}  
where ${\bf r}_{i}(t)$ denotes the position of point vortex $i$ at time $t$. Its
velocity  is given by
\begin{equation}
{\bf V}_{i}={d{\bf r}_{i}\over dt}=-{\bf z}\wedge\nabla\psi ({\bf r}={\bf
r}_{i},t),
\label{Vi}
\end{equation} 
where $\psi$ is a solution of the Poisson equation (\ref{poisson}) with the
vorticity field    (\ref{omegadelta}). In an unbounded fluid, one has
\begin{equation}
\psi({\bf r})=-{1\over 2\pi}\sum_{i=1}^{N}\gamma\ln |{\bf r}-{\bf r}_{i}|.
\label{psi}
\end{equation}  
Therefore, the velocity of a point vortex is equal to the sum of the velocities
${\bf V}(j\rightarrow i)$  produced by the  $N-1$ other vortices, i.e.
\begin{equation}
{\bf V}_{i}=\sum_{j\neq i}{\bf V}(j\rightarrow i)
\label{Vieq}
\end{equation} 
with  
\begin{equation}
{\bf V}(j\rightarrow i)=-{\gamma\over 2\pi}{\bf z}\wedge {{\bf r}_{j}-{\bf
r}_{i}\over |{\bf r}_{j}-{\bf r}_{i}|^{2}}.
\label{Vji}
\end{equation} 
The above dynamics can be cast in a Hamiltonian form \cite{kirchhoff}:
\begin{equation}
\gamma {d x_{i}\over dt}={\partial H\over\partial y_{i}},\qquad \gamma {d
y_{i}\over dt}=-{\partial H\over\partial x_{i}},
\label{kirchhoff}
\end{equation} 
\begin{equation}
H=-{1\over 4\pi}\sum_{i\neq j}\gamma^{2}\ln |{\bf r}_{i}-{\bf r}_{j}|,
\label{hamiltonian}
\end{equation} 
where the coordinates $(x,y)$ of the point vortices are canonically conjugate.
These equations of motion still apply when the fluid is restrained by
boundaries, in which case the Hamiltonian (\ref{hamiltonian}) is modified so as
to allow for vortex images, and may be constructed in terms of Green's functions
depending on the geometry of the domain. Since $H$ is not explicitly time
dependant, it is a constant of the motion and it represents the ``potential''
energy of the point vortices (we shall see later on that it also represents the
kinetic energy of the flow). Therefore, point vortices behave like particles in
interaction like electric charges or stars. Note, however, that the  Hamiltonian
(\ref{hamiltonian}) does not involve a ``kinetic'' energy of the vortices in the
usual sense. This is related to the particular circumstance that a point vortex
produces a velocity not an acceleration. As a result,  an isolated vortex
remains at rest contrary to a material particle which has a rectilinear motion
due to its inertia. Point vortices form therefore a very peculiar Hamiltonian
system.

\subsection{The microcanonical approach of Onsager}
\label{sec_onsager}

The statistical mechanics of point vortices was first considered by Onsager
\cite{onsager}  who showed the existence of negative temperatures at which point
vortices cluster into ``supervortices''. Let us briefly recall his
argumentation. 

Consider a liquid enclosed by a boundary, so that the vortices
are confined to an area $A$. Since the coordinates $(x,y)$ of the point vortices
are canonically conjugate, the phase space coincides with the configuration
space and is {\it finite}:
\begin{equation}
\int dx_{1}dy_{1}...dx_{N}dy_{N}=\biggl (\int dx dy\biggr )^{N}=A^{N}.
\label{AN}
\end{equation} 
This striking property contrasts with most classical Hamiltonian systems
considered in statistical mechanics which have unbounded phase spaces due to the
presence of a kinetic term in the Hamiltonian.

As is usual in the microcanonical description of a system of $N$ particles, we
introduce the density of state
\begin{equation}
g(E)=\int dx_{1}dy_{1}...dx_{N}dy_{N} \delta \biggl
(E-H(x_{1},y_{1},...,x_{N},y_{N})\biggr )
\label{gE}
\end{equation} 
which gives the phase space volume per unit interaction energy $E$. The phase
space volume which corresponds to energies $H$ less than a given value $E$ can
be written
\begin{equation}
\Phi(E)=\int_{E_{min}}^{E} g(E)dE.
\label{PhiE}
\end{equation}  
It increases monotonically from zero to $A^{N}$ when $E$ goes from $E_{min}$ to
$+\infty$. Therefore, $g(E)=d\Phi(E)/dE$ will have a maximum value at some
$E=E_{m}$, say, before decreasing to zero when $E\rightarrow +\infty$. 

In the microcanonical ensemble, the entropy and the temperature are defined by
\begin{equation}
S=\ln g(E) \qquad \beta={1\over T}={dS\over dE}.
\label{ST}
\end{equation} 
For $E>E_{m}$, $S(E)$ is a decreasing function of energy and consequently the
temperature is {\it negative}. Now, high energy states $E\gg E_{m}$ are clearly
those in which the vortices are crowded as close together as possible. For
energies only slighlty greater than $E_{m}$, the concentration will not be so
dramatic but there will be a tendency for the vortices to group themselves
together on a macroscopic scale and form ``clusters'' or ``supervortices''. By
contrast, for $E<E_{m}$, the temperature is positive and the vortices have the
tendency to accumulate on the boundary of the domain in order to decrease their
energy. For a system with positive and negative vortices, the negative
temperature states, achieved for relatively high energies, consist of two large
counter-rotating vortices physically well separated in the box. On the contrary
when $E\rightarrow -\infty$, the temperature is positive and vortices of
opposite circulation tend to pair off.

\subsection{The mean field approximation}
\label{sec_meanfield}

It is easy to show that the exact distribution of point vortices expressed in
terms of $\delta$-functions 
\begin{equation}
\omega_{exact} ({\bf r},t)=\sum_{i=1}^{N}\gamma\delta ({\bf r}-{\bf r}_{i}(t))
\label{omegaexact}
\end{equation}
is solution of the Euler equation
\begin{equation}
{\partial\omega_{ex}\over \partial t}+{\bf u}_{ex}\nabla\omega_{ex}=0,
\label{Eulerexact}
\end{equation}
where ${\bf u}_{ex}$ is the exact velocity field determined by equations
(\ref{div}) (\ref{poisson}) and (\ref{omegaexact}). This is proved as follows.
Taking the derivative of (\ref{omegaexact}) with respect to time, we obtain
\begin{eqnarray}
{\partial\omega_{ex}\over\partial t}=-\sum_{i=1}^{N}\gamma \nabla\delta({\bf
r}-{\bf r}_{i}(t)){\bf V}_{i}.
\label{De1}
\end{eqnarray}
Using ${\bf V}_{i}={\bf u}_{ex}({\bf r}_{i}(t),t)$, we can rewrite the foregoing
equation in the form
\begin{eqnarray}
{\partial\omega_{ex}\over\partial t}=-\nabla\sum_{i=1}^{N}\gamma \delta({\bf
r}-{\bf r}_{i}(t)){\bf u}_{ex}({\bf r},t).
\label{De2}
\end{eqnarray}
Since the velocity is divergenceless, we obtain 
\begin{eqnarray}
{\partial\omega_{ex}\over\partial t}({\bf r},t)=-{\bf u}_{ex}({\bf
r},t)\nabla\sum_{i=1}^{N}\gamma \delta({\bf r}-{\bf r}_{i}(t))=-{\bf
u}_{ex}\nabla\omega_{ex}({\bf r},t).
\label{De3}
\end{eqnarray}
Therefore, the Euler equation (\ref{Eulerexact}) with (\ref{div})
(\ref{poisson}) and (\ref{omegadelta}) contains exactly the same information as
the Hamiltonian system (\ref{kirchhoff})(\ref{hamiltonian}).

This description in terms of $\delta$-functions, while being technically
correct, is useless for practical purposes, because it requires the knowledge of
the exact trajectories of the point vortices for an arbitrary initial condition
(or the solution of the Euler equation (\ref{Eulerexact})). When $N$ is large,
this task is impossibly difficult. Therefore, instead of the exact vorticity
field expressed  in terms of delta functions, one is more interested by
functions which are smooth. For that reason, we introduce a smooth vorticity
field $\langle\omega\rangle({\bf r},t)$ which is proportional to the average
number of vortices contained in the cell $({\bf r},{\bf r}+d{\bf r})$ at time
$t$. This mean field description, which ignores the granularities of the sytem,
requires that it is possible to divide the domain  in a large number of cells in
such a way that each cell is (a) large enough to contain a macroscopic number of
point vortices but (b) small enough for all the particles in the cell can be
assumed to possess the same average characteristic of the cell.

In this mean field approximation, the Hamiltonian (\ref{hamiltonian}) is changed
into
\begin{equation}
E=-{1\over 4\pi}\int \langle\omega\rangle({\bf r})\langle\omega\rangle({\bf
r}')\ln |{\bf r}-{\bf r}'| d^{2}{\bf r} d^{2}{\bf r}'.
\label{E1}
\end{equation}
In writing this expression we have not taken into account the constraint $j\ne
i$ appearing in (\ref{hamiltonian}). Really, in (\ref{E1}) the integration
extends over the point ${\bf r}={\bf r}'$ so that (\ref{E1}) contains
self-energy terms which become infinitely large for point vortices. As will soon
become apparent, this meanfield approximation implies that the energy $E$ is
positive, a property which is not necessarily shared by the Hamiltonian
(\ref{hamiltonian}).

Using equation (\ref{psi}), adequatly generalized to account for a continuous
distribution of vortices, our expression (\ref{E1}) for $E$ can be rewritten
\begin{equation}
E={1\over 2}\int \langle\omega\rangle\psi  d^{2}{\bf r} .
\label{E2}
\end{equation}
Introducing explicitly the Poisson equation in (\ref{E2}) and integrating by
parts, one has successively
\begin{equation}
E={1\over 2}\int \psi (-\Delta\psi) d^{2}{\bf r}={1\over 2}\int (\nabla\psi)^{2}
d^{2}{\bf r} =\int {\langle{\bf u}\rangle^{2}\over 2}  d^{2}{\bf r}, 
\label{E3}
\end{equation}
where $\langle{\bf u}\rangle$ is the smooth velocity field (the second equality
is obtained by a part integration with the condition $\psi=0$ on the boundary).
Therefore, $E$ can be interpreted either as the potential energy of interaction
between vortices (see equation (\ref{E2})) or as the kinetic energy of the flow
(see equation (\ref{E3})).

\subsection{The mean field equilibrium}
\label{sec_montgomery} 

We now wish to determine the distribution of vortices at equilibrium following a
statistical mechanics approach \cite{jm}. To that purpose, using Boltzmann
procedure, we divide the macrocells $({\bf r},{\bf r}+d{\bf r})$ into a large
number of microcells and enumerate the number of ``microstates" which correspond
to the same ``macroscopic'' configuration of the system. The logarithm of this
number defines the entropy. In the mean field approximation, this leads to the
classical formula
\begin{equation}
S=-N\int P({\bf r})\ln P({\bf r}) d^{2}{\bf r},
\label{entropy}
\end{equation}
where $P({\bf r})$ is the density probability that a point vortex be in the
surface element centered on ${\bf r}$. The average vorticity in ${\bf r}$ is
related to this probability density by 
\begin{equation}
\langle\omega\rangle({\bf r})=N\gamma P({\bf r}).
\label{omegaaverage}
\end{equation}
At equilibrium, the system is in the most probable macroscopic state, i.e. the
state that is the most represented at the microscopic level. This optimal state
is obtained by maximizing the Boltzmann entropy (\ref{entropy}) at fixed energy
(\ref{E2}) and vortex number $N$, or total circulation 
\begin{equation}
\Gamma=N\gamma=\int\langle\omega\rangle d^{2}{\bf r}.
\label{circulation}
\end{equation}

Writing the variational principle in the form:
\begin{equation}
\delta S-\beta \delta E -\alpha\delta \Gamma=0,
\label{var}
\end{equation}
where $\beta$ and $\alpha$ are Lagrange multipliers, we find that the maximum
entropy state corresponds to the Boltzmann distribution \cite{jm}:
\begin{equation}
\langle \omega\rangle=Ae^{-\beta\gamma\psi}
\label{jm}
\end{equation}
with inverse temperature $\beta$. We can account for the conservation of angular
momentum  $L=\int \langle \omega \rangle r^{2} d^{2}{\bf r}$ (in a circular
domain) and impulse $P=\int \langle \omega \rangle y d^{2}{\bf r}$ (in a
channel) by introducing appropriate Lagrange multipliers $\Omega$ and $U$ for
each of these constraints. In that case equation (\ref{jm}) remains valid
provided that we replace the streamfunction $\psi$ by the relative
streamfunction $\psi'=\psi +{\Omega\over 2}r^{2}-U y$. This more general
situation has been considered in, e.g., Ref. \cite{classi} to describe rotating
or translating dipoles.

Substituting the Boltzmann relation between $\langle\omega\rangle$ and $\psi$ in
the Poisson equation (\ref{poisson}), we obtain a differential equation for  the
streamfunction:
\begin{equation}
-\Delta\psi=Ae^{-\beta\gamma\psi} \qquad (point\  vortices)
\label{pv}
\end{equation}
which determines the equilibrium distribution of vortices. In the case of
stellar systems and electric charges, the corresponding Boltzmann-Poisson
equation has the form \cite{pad}:
\begin{equation}
\Delta\Phi=4\pi G Ae^{-\beta' m\Phi} \qquad (stellar \ systems)
\label{stars}
\end{equation}
\begin{equation}
-\Delta\Phi= {A\over\epsilon_{0}} e^{-\beta' q\Phi}  \qquad (electric \ charges)
\label{charges}
\end{equation}
where $\Phi$ denotes successively the gravitational and the electrostatic
potential. For these systems, $\beta'>0$ since the temperature is a measure of
the kinetic energy. By contrast, for point vortices there is no kinetic term in
the Hamiltonian (\ref{hamiltonian}) and the temperature can be either positive
or negative. When $\beta<0$, equation (\ref{pv}) is similar in structure  to
equation (\ref{stars}). The vortices tend to attract each other, like stars in a
galaxy, and form ``clusters'' or ``supervortices''. The density profile
determined by (\ref{pv}) or (\ref{stars}) is a {\it decreasing} function of the
distance. When $\beta>0$, equation (\ref{pv}) is similar in structure to
equation (\ref{charges}). The vortices tend to repell each other, like electric
charges, and accumulate at the boundary. The density profile determined by
(\ref{pv}) or (\ref{charges}) is an {\it increasing} function of the distance.
Therefore, the formal analogy between 2D vortices and stellar systems is
intimately related to the existence of negative temperatures in 2D turbulence.
However, the physical  mechanism by which vortices and  stellar systems achieve
equilibrium is different. Whereas the organization of stars is relatively clear
because of the attractive nature of gravity, the organization of point vortices
at negative temperatures is much less intuitive. In the following section, we
shall give a physical interpretation of this phenomenon in terms of a
``systematic drift''.

\section{Elementary derivation of the systematic drift }
\label{sec_elementary}

\subsection{Analogy with Brownian motion: the necessity of the drift }
\label{sec_fp}

We shall first show the necessity of this drift by using an analogy with
Brownian theory. The starting point of this analogy is to realize that the
velocity of a point vortex can be decomposed in two terms: a smoothly varying
function of position and time $\langle {\bf V}\rangle({\bf r},t)$ and a function
$\mb{\cal V}(t)$  taking into account the ``granularity" of the system and
undergoing strong discontinuities. The total velocity of a point vortex can
therefore be written:
\begin{equation}
{\bf V}=\langle {\bf V}\rangle ({\bf r},t)+\mb{\cal V}(t).
\label{decomposition}
\end{equation}
The velocity $\langle {\bf V}\rangle ({\bf r},t)$ reflects the influence of the
system as a {\it whole} and is generated by the mean vorticity
$\langle\omega\rangle({\bf r},t)$ according to the Biot \& Savart formula:
\begin{equation}
\langle {\bf V}\rangle ({\bf r},t)=-{1\over 2\pi}{\bf z}\wedge \int  {{\bf
r}'-{\bf r}\over |{\bf r}'-{\bf r}|^{2}}\langle\omega\rangle({\bf
r}',t)d^{2}{\bf r}'.
\label{BiotSavart}
\end{equation} 
The fluctuation $\mb{\cal V}(t)$ arises from the difference between the exact
distribution of the point vortices $\omega_{exact}({\bf r},t)$ and their
``smoothed-out'' distribution $\langle\omega\rangle({\bf r},t)$. It is on
account of these fluctuations that the velocity of the test vortex will depart
from its mean field value $\langle {\bf V}\rangle$. The velocity fluctuation
$\cal V$, of order ${\gamma\over d}$ (where $d\sim n^{-1/2}$ is the inter-vortex
distance), is much smaller than the average velocity $\langle V\rangle$, of
order $n\gamma R$ (where $R$ is the domain size), but this term has a cumulative
effect which gives rise to a process of diffusion. It makes sense therefore to
introduce a stochastic description of the vortex motion like for colloidal
suspensions in a liquid \cite{kramers}  or stars in globular clusters
\cite{chandrasekhar}. However, contrary to the ideal Brownian motion, point
vortex systems have relatively long correlation times. This makes the study much
more complicated than usual and the technical study of section
\ref{sec_relaxation} is required. In order to gain some physical insight in the
problem, we shall ignore this difficulty for the moment and describe the system
by traditional stochastic processes.    

According to equation (\ref{decomposition}), we would naively expect that the
evolution of the density probability $P({\bf r},t)$ be governed by a diffusion
equation of the form
\begin{equation}
{\partial P\over \partial t}+\langle {\bf V}\rangle\nabla P=D\Delta P.
\label{diffusionA}
\end{equation} 
This would in fact be the case for a passive particle having no retroaction on
the vortices or when the distribution of vortices is uniform like in
\cite{weiss}\cite{cs}\cite{sc}. However, this diffusion equation cannot be valid
when the system is inhomogeneous. A first apparent reason is that equation
(\ref{diffusionA}) does not converge towards the Boltzmann distribution
(\ref{jm}) when $t\rightarrow +\infty$. Another related  difficulty is that
equation (\ref{diffusionA}) does not conserve energy. It seems therefore that a
term is missing to act against the diffusion.

These problems are similar to those encountered in Brownian theory or for
stellar systems. They have traditionally be solved by introducing a {\it
dynamical friction} in order to compensate for the effect of diffusion. The
occurence of this frictional force is a manifestation of the
``fluctuation-dissipation" theorem in statistical mechanics. In the present
context, the dynamical friction is replaced by a {\it systematic drift} of the
vortices. We must therefore rewrite the decomposition (\ref{decomposition}) in
the form
\begin{equation}
{\bf V}=\langle {\bf V}\rangle-\xi\nabla\psi+{\mb{\cal V}}(t),
\label{decomposition2}
\end{equation}
where $\xi$ is the drift coefficient. In section \ref{sec_polarization}, we
shall give a physical justification for the existence of the drift in terms of a
polarization process and in section \ref{sec_relaxation} we shall derive this
term directly from the Liouville equation by using projection operator technics.
The importance of this drift was first pointed out by Chavanis (1998) using a
thermal bath approximation and a linear response theory. The drift term must be
calculated by resorting to relatively elaborate technics but it is remarkable
that a general relationship between $\xi$ and $D$ can be obtained without beeing
required to analyze at any point  the details of the ``subdynamics''.

According to equation (\ref{decomposition2}), the equation of motion for a point
vortex can be written in the form
\begin{equation}
{d{\bf r}\over dt}=\langle {\bf V}\rangle-\xi\nabla\psi+{\mb{\cal V}}(t).
\label{motion}
\end{equation}
Since the velocity ${\mb{\cal V}}(t)$ undergoes strong discontinuities, the
trajectory ${\bf r}(t)$ of the point vortex is not differentiable. Therefore,
equation (\ref{motion}) must be viewed as a stochastic equation analogous to the
Langevin equation in the ordinary Brownian theory. Let $\Delta t$ be an interval
of time long compared to the fluctuation time but short at the scale on which
the physical parameters change appreciably. The variation in the position
$\Delta {\bf r}$ of the particle during $\Delta t$ is given by 
\begin{equation}
\Delta{\bf r}=\langle {\bf V}\rangle\Delta t-\xi\nabla\psi\Delta t+{\bf
B}(\Delta t),
\label{lang1}
\end{equation}
where 
\begin{equation}
{\bf B}(\Delta t)=\int_{t}^{t+\Delta t}{\mb {\cal V}}(t')dt'.
\label{B}
\end{equation}
Each fluctuation ${\mb{\cal V}}$ produces a small displacement $\delta {\bf r}$
but the repeated action of these fluctuations produces a net displacement  of
the same order as the drift $\xi\nabla\psi$. To determine the probability
$w[{\bf B}(\Delta t)]$ that the fluctuations produce a displacement ${\bf
B}(\Delta t)$ during the time interval $\Delta t$, we first divide the interval
$(t,t+\Delta t)$ into a succession of discrete increments in position and
observe that  ${\bf B}(\Delta t)$ is a sum of $N$ random variables $T({\cal
V}_{i}){\bf {\cal V}}_{i}$ where $T({\cal V})$ characterizes the typical
duration of the velocity fluctuation ${\mb {\cal V}}$. This is a problem of
random  walks where ${\bf B}(\Delta t)$ represents the distance reached after
$N$ steps. For large $N$'s, the Central Limit Theorem leads to a Gaussian
transition probability:
\begin{equation}
w[{\bf B}(\Delta t)]={1\over 4\pi D\Delta t}e^{-{B(\Delta t)^{2}\over 
4D\Delta t}}
\label{wB}
\end{equation} 
with a diffusion coefficient 
\begin{equation}
D={1\over 4}\langle T({\cal V}){\cal V}^{2}\rangle.
\label{Diffusioncoeff}
\end{equation}

We now assume that the motion of a point vortex can be idealized by a Markov
process, i.e. the probability at time $t+\Delta t$ depends on the probability at
time $t$ but not at earlier times. As indicated previously, this approximation
is not completely correct in the case of point vortices which have long
correlation times. However, relaxing this hypothesis would involve more
intricate equations (see section \ref{sec_relaxation}) and we shall ignore this
difficulty for the moment. We write therefore:
\begin{equation}
P({\bf r},t+\Delta t)=\int P({\bf r}-\Delta {\bf r},t)w({\bf r}-\Delta {\bf
r}|\Delta {\bf r})d^{2}(\Delta{\bf r}),
\label{markov}
\end{equation}
where $w({\bf r}-\Delta {\bf r}|\Delta {\bf r})$ is the probability for a point vortex located in ${\bf r}-\Delta{\bf r}$ to suffer an increment of position
 $\Delta {\bf r}$ during $\Delta t$. Expanding
$P({\bf r},t+\Delta t)$, $P({\bf r}-\Delta {\bf r},t)$ and 
$w({\bf r}-\Delta {\bf r}|\Delta {\bf r})$ in the form of Taylor
series, we arrive at the Fokker-Planck equation in its general form:
\begin{equation}
{\partial P\over\partial t}\Delta t=-\sum_{i}{\partial\over\partial
r_{i}}(P\langle \Delta r_{i}\rangle)+{1\over
2}\sum_{i,j}{\partial^{2}\over\partial r_{i}\partial r_{j}}(P\langle \Delta
r_{i}\Delta r_{j}\rangle),
\label{FP1}
\end{equation}
where
\begin{equation}
\langle \Delta r_{i}\rangle=\int \Delta r_{i}w({\bf r}|\Delta {\bf
r})d^{2}(\Delta{\bf r})
\label{M1}
\end{equation}
\begin{equation}
\langle \Delta r_{i} \Delta r_{j}\rangle=\int \Delta r_{i}\Delta r_{j}w({\bf
r}|\Delta {\bf r})d^{2}(\Delta{\bf r}).
\label{M2}
\end{equation} 
According to equation (\ref{wB}), the transition probability from ${\bf
r}$ to ${\bf r}+\Delta {\bf r}$ is given by
\begin{equation}
w({\bf r}|\Delta {\bf r})={1\over 4\pi D\Delta t}exp\biggl \lbrace
-{(\Delta {\bf r}-(\langle {\bf V}\rangle-\xi\nabla\psi)\Delta t)^{2}\over
4D\Delta t}\biggr\rbrace.
\label{transition}
\end{equation} 
With (\ref{transition}), the moments (\ref{M1}) (\ref{M2}) can be easily
evaluated yielding 
\begin{equation}
\langle \Delta {\bf r}\rangle=(\langle {\bf V}\rangle-\xi\nabla\psi)\Delta
t,\qquad \langle \Delta r_{i}\Delta r_{j}\rangle=2D\Delta t\delta_{ij}.
\label{M3}
\end{equation} 
Substituting these results in the general Fokker-Planck equation (\ref{FP1}), we
find that:
\begin{equation}
{\partial P\over\partial t}+\langle {\bf V}\rangle\nabla P=\nabla(D\nabla P+\xi
P\nabla\psi).
\label{FP2}
\end{equation} 
We had previously introduced this equation in Ref. \cite{chav98} using
phenomenological arguments. The physical interpretation of each term is
straightforward. The left hand side (which can be written $dP/dt$) is an
advection term due to the smooth mean field velocity $\langle {\bf V}\rangle$.
The right hand side can be written as the divergence of a current $-\nabla .
{\bf J}$ and is the sum of two terms: the first term is a diffusion due to the
erratic motion of the vortices caused by the fluctuations ${\mb{\cal V}}$; the
second term accounts for the systematic drift of the vortices. At equilibrium,
the drift precisely balances  random scatterings and the distribution (\ref{jm})
is settled. More precisely, the condition that the Maxwell-Boltzmann statistics
(\ref{jm}) satisfies equation (\ref{FP2}) identically requires that $D$ and
$\xi$ be related according to the relation
\begin{equation}
\xi= D\beta\gamma
\label{Einstein}
\end{equation} 
which is a  generalization of the Einstein formula to the case of point
vortices. A more rigorous justification of this relation will be given in
section \ref{sec_relaxation} where the diffusion coefficient and the drift term
are calculated explicitly.

\subsection{Systematic drift: the result of a polarization process}
\label{sec_polarization}

According to the previous discussion, the relaxation of a point vortex towards
statistical equilibrium can be described by a Fokker-Planck equation 
\begin{equation}
{\partial P\over\partial t}+\langle {\bf V}\rangle\nabla P=\nabla(D(\nabla 
P+\beta\gamma  P\nabla\psi))
\label{FP3}
\end{equation} 
involving a diffusion term $-D\nabla P$ and a drift term
\begin{equation}
\langle{\bf V}\rangle_{drift}=-D\beta\gamma \nabla\psi.
\label{drift}
\end{equation} 
The drift is normal to the mean field velocity $\langle {\bf V}\rangle=-{\bf
z}\wedge\nabla\psi$ of the vortices and its direction, depending on the sign of
$\beta$, has important physical implications. To fix the ideas, let us assume
that all point vortices have positive circulation (the opposite case leads to
the same conclusion). Due to the meanfield velocity, a particular point vortex
rotates anticlockwise. At negative temperatures, the drift is directed to its
left and the vortex is  {\it attracted} to the center of the domain. On the
contrary, at positive temperatures, the drift is directed to its right and the
vortex is {\it rejected} against the boundary. Therefore, the effect of the
drift is consistent with Onsager thermodynamical approach and it provides, in
addition, a physical mechanism to understand the organization of point vortices
at negative temperatures. For $\beta=0$, the medium is homogeneous and there is
no drift. Equation (\ref{FP3}) reduces to a pure diffusion equation like in Ref.
\cite{weiss,cs,sc}. Therefore, the drift occurs only in the presence of a
background shear.

In stellar systems, the relaxation of the distribution function $f({\bf r},{\bf
v},t)$ is usually described by the Kramers-Chandrasekhar equation: 
\begin{equation}
{\partial f\over\partial t}+{\bf v}{\partial f\over \partial {\bf r}}+\langle
{\bf F}\rangle {\partial f\over\partial {\bf v}}={\partial\over \partial{\bf
v}}\biggl \lbrace D\biggl ({\partial f\over\partial {\bf v}}+\beta  m f {\bf
v}\biggr) \biggr\rbrace 
\label{KC}
\end{equation}
which is a particular Fokker-Planck equation with a structure analogous to
equation (\ref{FP3}). In this analogy, the dynamical friction experienced by a
star as a result of close encounters (Chandrasekhar \cite{chandrasekhar}):   
\begin{equation}
\langle{\bf F}\rangle_{friction}=-D\beta m  {\bf v}
\label{friction}
\end{equation} 
is the counterpart of the systematic drift (\ref{drift}) experienced by a point
vortex in two dimensional turbulence. The dynamical friction can be viewed as
the drag exerted on a test star by the wake it induces in the field stars, like
in a polarization process. We can use a similar approach to understand the
origin of the drift. Let us consider a collection of $N$ point vortices at
statistical equilibrium with inverse temperature $\beta$. When $\beta<0$, the
density of these ``field'' vortices decreases from the center to the periphery
of the domain. A ``test'' vortex moving through this medium  locally modifies
the vorticity distribution and produces a polarization cloud which amounts
typically to a rotation of the surrounding vortices. This creates an excess of
density behind it an a deficit of density in front of it. Therefore, the
retroaction of the field vortices leads to a drift of the test vortex directed
inward. We reach the opposite conclusion if the temperature is positive. When
the system is homogeneous ($\beta=0$ in a domain with no special symmetry), the
polarization cloud induced by the test vortex has no effect and the drift
cancels out. Therefore, a homogeneous system of point vortices remains
homogeneous \cite{weiss,cs,sc}.

\subsection{The Maximum Entropy Production Principle }
\label{sec_mepp}

In section \ref{sec_relaxation}, we shall derive the Fokker-Planck equation
(\ref{FP3}) directly from the Liouville equation. However, we want to show first
that the general structure of this equation can be understood from relatively
simple thermodynamical arguments.

Let us  rewrite equation (\ref{FP3}) in the form  
\begin{equation}
{\partial\langle \omega\rangle\over \partial t}+\langle {\bf
u}\rangle\nabla\langle\omega\rangle=-\nabla .{\bf J},
\label{diffomega}
\end{equation}
where ${\bf J}$ is an unknown diffusion current. This equation conserves the
circulation (\ref{circulation}) provided that ${\bf J}. {\hat {\bf n}}=0$ on the
domain boundary (${\hat {\bf n}}$ is a unit vector normal to the boundary). The
problem at hands consists in determining the expression for ${\bf J}$. Its exact
expression depends on microscopic processes and is therefore difficult to
capture. However, it is easy to write down some macroscopic constraints that it
must satisfy. These constraints are provided by the first and second principles
of thermodynamics, namely the conservation of energy and the increase of
entropy. We shall find that these constraints are very stringent and determine
completely the structure of the diffusion current. 

Taking the time derivative of equations (\ref{E2}) (\ref{entropy}) and
substituting for (\ref{diffomega}) we obtain the constraints
\begin{equation}
\dot E=\int {\bf J}.\nabla\psi d^{2}{\bf r}=0 
\label{dotE}
\end{equation}
and 
\begin{equation}
\dot S=-{1\over\gamma} \int {\bf J}.\nabla\ln \langle \omega\rangle d^{2}{\bf
r}\ge 0.
\label{dotS}
\end{equation}
We shall now introduce an optimization procedure known as the Maximum Entropy
Production Principle (M.E.P.P.). This principle was introduced initially in the
context of 2D turbulence by Robert \& Sommeria \cite{rs} but its domain of
applicability is very general and concerns, for example, the case of stellar
systems\cite{csr,chavmepp}. This principle states that  ``out of equilibrium,
the system evolves so as to maximize its rate of entropy production $\dot S$
while accounting for all the constraints imposed by the dynamics, in particular
the conservation of energy $\dot E=0$''. There is no precise justification for
this principle and it is important therefore to confront the M.E.P.P. with more
rigorous methods, like those of section \ref{sec_relaxation}, to determine its
domain of validity. In any case, the M.E.P.P. can be considered  as a convenient
tool to build relaxation equations which are mathematically well-behaved and
which can serve as numerical algorithms to calculate maximum entropy states.

We seek therefore the optimal diffusion current ${\bf J}$ which maximizes the
rate of entropy production $\dot S$ at fixed energy. In order to avoid the
unphysical solution $|{\bf J}|\rightarrow +\infty$ with ${\bf J}.\nabla\psi=0$,
we impose the additional constraint 
\begin{equation}
J^{2}\le C({\bf r},t),
\label{bound}
\end{equation}
where $C$ is an upper bound which must exist but is not known. The solution of the optimization problem is
\begin{equation}
{\bf
J}=-D(\nabla\langle\omega\rangle+\beta(t)\gamma\langle\omega\rangle\nabla\psi),
\label{Jopt}
\end{equation}
where $\beta(t)$ and $D({\bf r},t)$ are Lagrange multipliers associated with the
constraints (\ref{dotE}) and (\ref{bound}). When substituted in equation
(\ref{diffomega}), we obtain:
\begin{equation}
{\partial \langle\omega\rangle\over\partial t}+\langle {\bf V}\rangle\nabla
\langle\omega\rangle=\nabla (D(\nabla \langle\omega\rangle+\beta(t) \gamma
\langle\omega\rangle \nabla\psi)). 
\label{FPmepp}
\end{equation}
This equation has the same form as the Fokker-Planck equation (\ref{FP3}). Here,
the diffusion term arises from the variations of entropy $\delta \dot S$ and the
drift term is necessary to conserve energy. Note that the Einstein formula
(\ref{Einstein}) is automatically satisfied by this variational approach. 

The time evolution of the inverse temperature $\beta(t)$ is determined by the
conservation of energy. Substituting the diffusion current (\ref{Jopt}) in the
constraint (\ref{dotE}) we find 
\begin{equation}
\beta(t)=-{\int D\nabla\langle\omega\rangle . \nabla\psi d^{2}{\bf r}\over\int
D\gamma \langle\omega\rangle (\nabla\psi)^{2}d^{2}{\bf r}}.
\label{betat}
\end{equation}
We can also check that the entropy monotonically increases during the relaxation
provided that $D\ge 0$. Indeed, using (\ref{dotS})(\ref{Jopt}) and (\ref{dotE})
we can easily establish that
\begin{equation}
\dot S=\int {J^{2}\over D\gamma \langle\omega\rangle}d^{2}{\bf r}\ge 0.
\label{Hteo}
\end{equation}
At equilibrium, ${\bf J}=0$ and we recover the Boltzmann distribution
(\ref{jm}).

Note that the optimal current (\ref{Jopt}) can be written ${\bf
J}=\chi\nabla\alpha$ where $\alpha=\ln\langle \omega\rangle+\beta\gamma\psi$ is
a ``generalized potential'' which is uniform at equilibrium. Therefore, the
M.E.P.P. can be viewed as a variational formulation of the linear thermodynamics
of Onsager which relates the diffusion currents to the gradients of generalized
potentials.  However, the M.E.P.P. gives a more elegant approach to the problem
and, conceding a real importance to the constraints, it is easier to implement
in more complicated situations \cite{rs,csr,cs97}. In addition, it shows that
the structure of the relaxation is determined by purely thermodynamical
arguments. All explicit reference to the subdynamics is encapsulated in the
diffusion coefficient which is left unspecified (it appears as a  Lagrange
multiplier related to an unknown bound on the diffusion current).  It must be
therefore calculated with a more microscopic model like the one of section
\ref{sec_relaxation}.

\section{Relaxation of a point vortex in a thermal bath}
\label{sec_relaxation}

\subsection{The Liouville equation}
\label{sec_liouville}

Let us consider a collection of $N+1$ point vortices with identical circulation
$\gamma$. We select one of these vortices, for example point vortex $0$, and
call it the ``test vortex''. The other vortices $1,...,N$ will be refered to as
the ``field vortices''. Let $\mu({\bf r},{\bf r}_{1},...,{\bf r}_{N},t)$ denote
the $N+1$ particle distribution of the system, i.e $\mu({\bf r},{\bf
r}_{1},...,{\bf r}_{N},t)d^{2}{\bf r}d^{2}{\bf r}_{1}...d^{2}{\bf r}_{N}$
represents the probability that point vortex $0$ be in the cell $({\bf r},{\bf
r}+d{\bf r})$, point vortex $1$ in the cell $({\bf r}_{1},{\bf r}_{1}+d{\bf
r}_{1})$... and point vortex $N$ in the cell $({\bf r}_{N},{\bf r}_{N}+d{\bf
r}_{N})$ at time $t$. The $(N+1)$-particle distribution function $\mu(t)$
satisfies the Liouville equation
\begin{equation}
{\partial\mu\over\partial t}+\sum_{i=0}^{N}{\bf V}_{i}{\partial\mu\over\partial
{\bf r}_{i}}=0,
\label{Liouville1}
\end{equation}
where ${\bf V}_{i}$ is the velocity of vortex $i$ produced by the other vortices
according to equations (\ref{Vieq}) (\ref{Vji}). We also introduce the one- and
$N$-particle distribution functions defined by
\begin{equation}
P({\bf r},t)=\int \mu(\lbrace{\bf r}_{k}\rbrace,t) \prod_{k=1}^{N}d^{2}{\bf
r}_{k},
\label{P0}
\end{equation}
\begin{equation}
\mu_{sys}({\bf r}_{1},...,{\bf r}_{N},t)=\int \mu(\lbrace{\bf
r}_{k}\rbrace,t)d^{2}{\bf r}.
\label{musys}
\end{equation}
The physical picture that we have in mind is that the test vortex evolves in a
``bath'' of field vortices. Therefore, we rewrite the distribution function
$\mu$ in the suggestive form
\begin{equation}
\mu({\bf r},{\bf r}_{1},...,{\bf r}_{N},t)=P({\bf r},t)\mu_{sys}({\bf
r}_{1},...,{\bf r}_{N},t)+\mu_{I}({\bf r},{\bf r}_{1},...,{\bf r}_{N},t),
\label{muI}
\end{equation}
where $\mu_{I}$ reflects the effect of correlations between the test vortex and
the field vortices. Physically, this term accounts for the polarization process
described qualitatively in section \ref{sec_polarization}.

The Liouville equation (\ref{Liouville1}) provides the correct starting point
for the analysis of the dynamics of our vortex system. However, when $N$ is
large, this equation contains much more information than one can interpret.
Consequently, what one would like to do is to describe the system in some
average sense by a one-particle distribution function. In the previous sections
we have indeed derived heuristically some differential equation satisfied by
this distribution function on the basis of stochastic arguments. We shall now
discuss the connection of such heuristic theories with a more microscopic
description of the system.

\subsection{The projection operator formalism}
\label{sec_projection}

Our first objective is to derive some {\it exact} kinetic equations satisfied by
$P({\bf r},t)$ and $\mu_{sys}({\bf r}_{1},...,{\bf r}_{N},t)$. This can be
achieved by using the  projection operator formalism developed by Willis \&
Picard \cite{wp}. This formalism was also used by Kandrup \cite{kandrup} in the
context of stellar dynamics to derive a generalized Landau equation describing
the time evolution of the distribution function of stars in an inhomogeneous
medium. We shall just recall the main steps of the theory. More details can be
found in the original paper  of  Willis \& Picard \cite{wp} and in Kandrup
\cite{kandrup}. To have similar notations, we set $x\equiv \lbrace{\bf
r}\rbrace$ and $y\equiv \lbrace {\bf r}_{1},...,{\bf r}_{N}\rbrace$. Then,
equation (\ref{muI}) can be put in the form  
\begin{equation}
\mu(x,y,t)=\mu_{R}(x,y,t)+\mu_{I}(x,y,t)
\label{muIbis}
\end{equation}
with
\begin{equation}
\mu_{R}(x,y,t)=f(x,t)g(y,t),
\label{muR}
\end{equation}
where we have written $f(x,t)\equiv P({\bf r},t)$ and $g(y,t)\equiv
\mu_{sys}({\bf r}_{1},...,{\bf r}_{N},t)$. The Liouville equation is also cast
in the form
\begin{equation}
{\partial\mu\over\partial t}=-i L\mu=-i(L_{0}+L_{sys}+L')\mu,
\label{Liouville2}
\end{equation}
where $L_{0}$ and $L_{sys}$ act respectively only on the variables $x$ and $y$,
whereas the interaction Liouvillian $L'$ acts upon both $x$ and $y$ (the complex
number $i$ is here purely formal and has been introduced only to have the same
notations as Ref. \cite{wp,kandrup}).

Following Willis \& Picard, we introduce the time-dependant projection operator:
\begin{equation}
{ P}(x,y,t)=g(y,t)\int dy+f(x,t)\int dx-f(x,t)g(y,t)\int dx\int dy.
\label{Proj}
\end{equation}
We can easily check that:
\begin{equation}
{ P}(x,y,t)\mu(x,y,t)=\mu_{R}(x,y,t),
\label{Pmu}
\end{equation}
\begin{equation}
\lbrack 1-{ P}(x,y,t)\rbrack \mu(x,y,t)=\mu_{I}(x,y,t).
\label{Perpmu}
\end{equation}
We also verify that ${ P}$ is a projection in the sense that ${ P}^{2}(t)={
P}(t)$. Applying ${ P}$ and $1-{ P}$ on the Liouville equation
(\ref{Liouville2}), we obtain the coupled equations
\begin{equation}
\partial_{t}\mu_{R}(x,y,t)=-i{ P}L\mu_{R}-iPL\mu_{I}
\label{Sys1}
\end{equation}
and 
\begin{equation}
\partial_{t}\mu_{I}(x,y,t)=-i(1-P)L\mu_{R}-i(1-P)L\mu_{I}.
\label{Sys2}
\end{equation}
These equations should be compared with equations (8) (9) that appear in the
quasilinear theory of 2D turbulence \cite{chavquasi}. In the present context,
equations (\ref{Sys1}) (\ref{Sys2}) describe the separation between a
``macrodynamics'' and a ``subdynamics''. In the quasilinear theory, equations
(8) (9) describe the evolution of the ``coarse-grained'' and  ``fine-grained''
components of the vorticity.

Introducing the Greenian
\begin{equation}
{\cal G}(t,t')\equiv \exp \Biggl \lbrace -i\int_{t'}^{t}dt'' \lbrack 
1-{ P}(t'')\rbrack{L}\Biggr \rbrace,
\label{Greenian}
\end{equation}
we can immediately write down a formal solution of equation (\ref{Sys2}),
namely:
\begin{equation}
\mu_{I}(x,y,t)=-\int_{0}^{t}dt' {\cal G}(t,t')i\lbrack 1-{P}(t')\rbrack { L}\mu_{R}(x,y,t'),
\label{muIexp}
\end{equation}
where we have assumed that initially the particles are uncorrelated so that
$\mu_{I}(x,y,0)=0$. Substituting for $\mu_{I}(x,y,t)$ from equation
(\ref{muIexp}) in equation (\ref{Sys1}), we obtain 
\begin{equation}
\partial_{t}\mu_{R}(x,y,t)=-i{ P}L\mu_{R}-\int_{0}^{t}dt' P(t)L {\cal G}(t,t')
\lbrack 1-P(t')\rbrack { L}\mu_{R}(x,y,t').
\label{sys1}
\end{equation}
The integration over $y$ will yield an equation describing the evolution of $f$.
Using some mathematical properties of the projection operator (\ref{Proj}), the
final result can be put in the nice symmetrical form given by Willis \& Picard
\cite{wp}:
\begin{equation}
\partial_{t}f(x,t)+i L_{0}f+i\langle L'\rangle_{sys}f=-\int_{0}^{t}dt'\int
dy\Delta_{t}L'{\cal G}(t,t')\Delta_{t'}L' g(y,t')f(x,t'),
\label{evolf}
\end{equation}
where the notations stand for
\begin{equation}
\langle L'\rangle_{sys}=\int dy' L'(x,y') g(y',t),
\label{Lprime}
\end{equation}
\begin{equation}
\langle L'\rangle_{0}=\int dx' L'(x',y) f(x',t),
\label{Lprime0}
\end{equation}
\begin{equation}
\Delta_{t}L'=L'-\langle L'\rangle_{sys}-\langle L'\rangle_{0}.
\label{DeltaLprime}
\end{equation}
Similarly, after integrating over $x$ we find the equation satisfied by $g$
\cite{wp}: 
\begin{equation}
\partial_{t}g(y,t)+i L_{sys}g-i\langle L'\rangle_{1}g=-\int_{0}^{t}dt'\int
dx\Delta_{t}L'{\cal G}(t,t')\Delta_{t'}L' g(y,t')f(x,t').
\label{evolg}
\end{equation}

\subsection{Application to the point vortex system}
\label{sec_application}

The previous theory is completely general and we now consider its application to
a system of point vortices. Let us first rewrite the Liouville equation
(\ref{Liouville1}) in a form that separates the contribution of the test vortex
from the contribution of the field vortices:
\begin{eqnarray}
{\partial\mu\over\partial t}+\sum_{i=1}^{N}{\bf V}(i\rightarrow
0){\partial\mu\over\partial {\bf r}}+\sum_{i=1}^{N}{\bf V}(0\rightarrow
i){\partial\mu\over\partial {\bf r}_{i}}\nonumber\\
+\sum_{i=1}^{N}\sum_{j=1,j\neq i}^{N}{\bf V}(j\rightarrow
i){\partial\mu\over\partial {\bf r}_{i}}=0.
\label{Liouvillesep}
\end{eqnarray}
The different operators that arise in the decomposition (\ref{Liouville2}) are
\begin{equation}
i L_{0}=0,
\label{L0}
\end{equation}
\begin{equation}
i L_{sys}=\sum_{i=1}^{N}\sum_{j\neq i,0}{\bf V}(j\rightarrow
i){\partial\over\partial {\bf r}_{i}},
\label{Lsys}
\end{equation}
\begin{equation}
i L'=\sum_{i=1}^{N}\biggl\lbrace {\bf V}(i\rightarrow 0){\partial\over\partial
{\bf r}}+{\bf V}(0\rightarrow i){\partial\over\partial {\bf r}_{i}}
\biggr\rbrace.
\label{Lprimebis}
\end{equation}

The mean-field velocity created by the field vortex $i$ on the test vortex is
denoted by
\begin{equation}
\langle {\bf V}(i\rightarrow 0)\rangle=\int P({\bf r}_{i},t) {\bf
V}(i\rightarrow 0) d^{2}{\bf r}_{i}.
\label{Vi0moy}
\end{equation}
Similarly, 
\begin{equation}
\langle {\bf V}(0\rightarrow i)\rangle=\int P({\bf r},t) {\bf V}(0\rightarrow i)
d^{2}{\bf r}
\label{V0imoy}
\end{equation}
denotes the mean field velocity created by the test vortex on the field vortex
$i$. Finally, the total mean field velocity experienced by the test vortex is
given by
\begin{eqnarray}
\langle {\bf V}\rangle=\sum_{i=1}^{N}\langle {\bf V}(i\rightarrow
0)\rangle=\sum_{i=1}^{N}\langle {\bf V}(1\rightarrow 0)\rangle=N\langle {\bf
V}(1\rightarrow 0)\rangle,
\label{Vmoy}
\end{eqnarray}
where the second equality follows from the identity of the point vortices.

We are now ready to evaluate the quantities (\ref{Lprime}) (\ref{Lprime0})
(\ref{DeltaLprime}). After straightforward integration by parts, we find
successively
\begin{equation}
i \langle L'\rangle_{sys}=\langle {\bf V}\rangle {\partial\over\partial {\bf
r}}=\sum_{i=1}^{N}\langle {\bf V}(i\rightarrow 0)\rangle {\partial\over\partial
{\bf r}},
\label{Q1}
\end{equation}
\begin{equation}
i \langle L'\rangle_{0}=\sum_{i=1}^{N}\langle {\bf V}(0\rightarrow
i)\rangle{\partial\over\partial {\bf r}_{i}},
\label{Q2}
\end{equation}
\begin{equation}
i \Delta_{t}L'=\sum_{i=1}^{N}\biggl\lbrace {\bf V}(i\rightarrow 0)-\langle{\bf
V}(i\rightarrow 0)\rangle\biggr \rbrace {\partial\over\partial {\bf
r}}+\sum_{i=1}^{N}\biggl\lbrace {\bf V}(0\rightarrow i)-\langle{\bf
V}(0\rightarrow i)\rangle\biggr\rbrace {\partial\over\partial {\bf r}_{i}}.
\label{Q3}
\end{equation}
Introducing the velocity fluctuations:
\begin{equation}
{\mb{\cal V}}(i\rightarrow 0)={\bf V}(i\rightarrow 0)-\langle{\bf
V}(i\rightarrow 0)\rangle,
\label{Vfluc1}
\end{equation}
\begin{equation}
{\mb{\cal V}}(0\rightarrow i)={\bf V}(0\rightarrow i)-\langle{\bf
V}(0\rightarrow i)\rangle,
\label{Vfluc2}
\end{equation}
we can rewrite our expression for $\Delta_{t}L'$ in the form
\begin{equation}
i\Delta_{t}L'=\sum_{i=1}^{N} {\mb{\cal V}}(i\rightarrow 0)
{\partial\over\partial {\bf r}}+\sum_{i=1}^{N}{\mb{\cal V}}(0\rightarrow i)
{\partial\over\partial {\bf r}_{i}}.
\label{Q4}
\end{equation}

Substituting these results in equation (\ref{evolf}), we obtain the following
kinetic equation for the one-particule distribution function  of a vortex
system:
\begin{eqnarray}
{\partial P\over\partial t}+\langle {\bf V}\rangle{\partial P\over\partial {\bf
r}}=\int_{0}^{t}dt'\int\prod_{k=1}^{N}d^{2}{\bf r}_{k}\sum_{i=1}^{N}{\mb{\cal
V}}(i\rightarrow 0){\partial\over\partial {\bf r}}\nonumber\\ 
\times{\cal
G}(t,t')\biggl \lbrace \sum_{j=1}^{N}{\mb{\cal V}}(j\rightarrow
0){\partial\over\partial {\bf r}}+\sum_{j=1}^{N}{\mb{\cal V}}(0\rightarrow
j){\partial\over\partial {\bf r}_{j}}\biggr\rbrace P({\bf
r},t')\mu_{sys}(\lbrace {\bf r}_{k}\rbrace,t')
\label{kin1}
\end{eqnarray}
or, alternatively,
\begin{eqnarray}
{\partial P\over\partial t}+\langle {\bf V}\rangle{\partial P\over\partial {\bf
r}}={\partial\over\partial r^{\mu}}\int_{0}^{t}d\tau\int\prod_{k=1}^{N}d^{2}{\bf
r}_{k}\sum_{i=1}^{N}\sum_{j=1}^{N}  {\cal V}^{\mu}(i\rightarrow 0)\nonumber\\
\times{\cal G}(t,t-\tau)\biggl ({\cal V}^{\nu}(j\rightarrow
0){\partial\over\partial {r}^{\nu}}+{{\cal V}}^{\nu}(0\rightarrow
j){\partial\over\partial { r}_{j}^{\nu}}\biggr ) P({\bf
r},t-\tau)\mu_{sys}(\lbrace {\bf r}_{k}\rbrace,t-\tau),
\label{kin2}
\end{eqnarray}
where the Greek indices refer to the  components of ${\mb{\cal V}}$ in a fixed
system of coordinates. We can note that equation (\ref{kin2}) already shares
some analogies with the Fokker-Planck equation of section \ref{sec_elementary}.
Indeed, the first term on the r.h.s. corresponds to a diffusion and the second
term to a drift. For a passive particle ${\cal V}^{\nu}(0\rightarrow j)=0$ and
the drift cancels out, as expected. 

\subsection{The thermal bath approximation}
\label{sec_bath}

Equation (\ref{kin2}) is an {\it exact} differential equation for $P({\bf
r},t)$. However, this equation is not directly soluble since the unkown function
$\mu_{sys}(\lbrace {\bf r}_{k}\rbrace,t)$ is given by an equation of the form
(\ref{evolg}) depending in turn on $P({\bf r},t)$. We therefore have to solve
the coupled system (\ref{evolf})-(\ref{evolg}). This system bears exactly the
same information as the initial Liouville equation (\ref{Liouville2})  and,
without further simplification, is untractable.

To reduce the complexity of the problem, we shall  implement a ``thermal bath
approximation''. We assume that the field vortices are in statistical
equilibrium with inverse temperature $\beta_{eq}$. Therefore, the $N$-particle
distribution function $\mu_{sys}(\lbrace {\bf r}_{k}\rbrace)$ can be
approximated by a product of $N$ one-particle distribution functions $P_{eq}$
given by the Maxwell-Boltzmann statistics (\ref{jm}). In other words, we make
the approximation
\begin{equation}
\mu_{sys}({\bf r}_{1},...,{\bf r}_{N},t)\simeq \mu_{eq}({\bf r}_{1},...,{\bf
r}_{N})
\label{thermalbath}
\end{equation}
with
\begin{equation}
\mu_{eq}({\bf r}_{1},...,{\bf r}_{N})=\prod_{k=1}^{N} P_{eq}({\bf
r}_{k})=\prod_{k=1}^{N}A_{k}e^{-\beta_{eq}\gamma\psi_{eq}({\bf r}_{k})},
\label{factorization}
\end{equation}
where $\psi_{eq}$ is solution of the Poisson equation (\ref{poisson}) with the
equilibrium vorticity $\langle\omega\rangle_{eq}=N\gamma P_{eq}$. Substituting
explicitly for the Boltzmann distribution from equation (\ref{factorization}) in
equation (\ref{kin2}), we obtain
\begin{eqnarray}
{\partial P\over\partial t}+\langle {\bf V}\rangle_{eq} {\partial P\over\partial
{\bf r}}={\partial\over\partial
r^{\mu}}\int_{0}^{t}d\tau\int\prod_{k=1}^{N}d^{2}{\bf
r}_{k}\sum_{i=1}^{N}\sum_{j=1}^{N}  {\cal V}^{\mu}(i\rightarrow 0)\nonumber\\
\times{\cal G}(t,t-\tau)\biggl ({\cal V}^{\nu}(j\rightarrow
0){\partial\over\partial {r}^{\nu}}-\beta_{eq}\gamma{{\cal
V}}^{\nu}(0\rightarrow j) {\partial\psi_{eq}\over\partial { r}_{j}^{\nu}}\biggr
) P({\bf r},t-\tau)\prod_{k=1}^{N}P_{eq}({\bf r}_{k}),
\label{K1}
\end{eqnarray}
where $\langle .\rangle_{eq}$ denotes the average with respect to the
equilibriun distribution $P_{eq}$. Explicating the action of the Greenian, we
can rewrite our equation (\ref{K1}) in the form
\begin{eqnarray}
{\partial P\over\partial t}+\langle {\bf V}\rangle_{eq}{\partial P\over\partial
{\bf r}}={\partial\over\partial
r^{\mu}}\int_{0}^{t}d\tau\int\prod_{k=1}^{N}d^{2}{\bf
r}_{k}\sum_{i=1}^{N}\sum_{j=1}^{N}  {\cal V}^{\mu}(i\rightarrow 0,t)\nonumber\\
\times\biggl ({\cal V}^{\nu}(j\rightarrow 0,t-\tau){\partial\over\partial
{r}^{\nu}}-\beta_{eq}\gamma{{\cal V}}^{\nu}(0\rightarrow
j,t-\tau){\partial\psi_{eq}\over\partial { r}_{j}^{\nu}}({\bf
r}_{j}(t-\tau))\biggr ) \nonumber\\ 
\times P({\bf
r}(t-\tau),t-\tau)\prod_{k=1}^{N}P_{eq}({\bf r}_{k}(t-\tau)),
\label{K2}
\end{eqnarray}
where the retarded velocity ${\mb {\cal V}}(j\rightarrow i,t-\tau)$ must be
viewed as an explicit function of time. More precisely ${\mb {\cal
V}}(j\rightarrow i,t-\tau)$ is a shorthand notation for ${\mb {\cal V}}({\bf
r}_{j}(t-\tau)\rightarrow {\bf r}_{i}(t-\tau))$, where ${\bf r}_{i}(t-\tau)$
denotes the position at time $t-\tau$ of the $i$-th point vortex located in
${\bf r}_{i}\equiv {\bf r}_{i}(t)$ at time $t$. The trajectories of the point
vortices between $t-\tau$ and $t$ are determined by the complicated Greenian
${\cal G}(t,t-\tau)$ defined in (\ref{Greenian}). We need therefore to solve the
exact Kirchhoff-Hamilton equations of motion (\ref{kirchhoff}). In fact, to a
good approximation, we can consider that the point vortices are purely advected
by the equilibrium  mean field velocity $\langle{\bf V}\rangle_{eq}$. Indeed,
when $N\rightarrow \infty$, we have already indicated that the velocity
fluctuation ${\bf {\cal V}}$ is much smaller than the mean field velocity
$\langle {\bf V}\rangle_{eq}$. Therefore, we can replace the exact Greenian
${\cal G}$ by a smoother Greenian $\langle {\cal G}\rangle_{eq}$ which would be
obtained if the point vortices were moving in the velocity field created by the
equilibrium distribution function $\mu_{eq}$. Formally, this Greenian is
constructed with the averaged Liouville operator $\langle{
L}\rangle_{eq}\equiv \sum_{i=0}^{N} \langle{\bf V}_{i}\rangle_{eq} {\partial
\over \partial {{\bf r}}_{i}}$. 

In this approximation, the correlations involving two different vortex pairs
vanish and the equation can be simplified considerably. Using the results of
Appendix \ref{sec_diffcoeff} (see in particular equation (\ref{Acorr2})), we
find
\begin{eqnarray}
{\partial P\over \partial t}+\langle {\bf V}\rangle_{eq}{\partial P\over\partial
{\bf r}}=\sum_{i=1}^{N} {\partial\over\partial r^{\mu}}\int_{0}^{t}d\tau \int
d^{2}{\bf r}_{i}{ V}^{\mu}(i\rightarrow 0,t)\nonumber\\ 
\times\biggl ({
V}^{\nu}(i\rightarrow 0, t-\tau){\partial \over\partial
r^{\nu}}-\beta_{eq}\gamma { V}^{\nu}(0\rightarrow i,t-\tau)   {\partial
\psi_{eq}\over\partial r_{i}^{\nu}}({\bf r}_{i}(t-\tau))\biggr )\nonumber\\
\times P({\bf r}(t-\tau),t-\tau)P_{eq}({\bf r}_{i}),
\label{K3}
\end{eqnarray}    
where we have used $P_{eq}({\bf r}_{i}(t-\tau))=P_{eq}({\bf r}_{i})$ since
$P_{eq}=f(\psi_{eq})$ is constant along a streamline and the particles are
assumed to follow the streamlines in a first approximation. Since the vortices
are identical we also have
\begin{eqnarray}
{\partial P\over \partial t}+\langle {\bf V}\rangle_{eq}{\partial P\over\partial
{\bf r}}=N {\partial\over\partial r^{\mu}}\int_{0}^{t}d\tau\int d^{2}{\bf
r}_{1}{ V}^{\mu}(1\rightarrow 0,t)\nonumber\\ 
\times\biggl ({
V}^{\nu}(1\rightarrow 0, t-\tau){\partial \over\partial
r^{\nu}}-\beta_{eq}\gamma { V}^{\nu}(0\rightarrow 1,t-\tau)   {\partial
\psi_{eq}\over\partial r_{1}^{\nu}}({\bf r}_{1}(t-\tau))\biggr )\nonumber\\
\times P({\bf r}(t-\tau),t-\tau)P_{eq}({\bf r}_{1}).
\label{K4}
\end{eqnarray}
Noting that the integral is dominated by the divergence of the product
$V^{\mu}V^{\nu}$ when ${\bf r}_{1}\rightarrow {\bf r}$, we can make a ``local
approximation'' and replace  ${\partial\psi_{eq}\over\partial r_{1}^{\nu}}({\bf
r}_{1})$   and $P_{eq}({\bf r}_{1})$ by their values taken in ${\bf r}$. For the
same reason, we can neglect vortex images and replace the Kernel ${\bf
V}(0\rightarrow 1)$ by its singular part (\ref{Vji}) satisfying  ${\bf
V}(0\rightarrow 1)=-{\bf V}(0\rightarrow 1)$. With these approximations, the
kinetic equation (\ref{K4}) takes the form:  
\begin{eqnarray}
{\partial P\over \partial t}+\langle {\bf V}\rangle_{eq}{\partial P\over\partial
{\bf r}}= N {\partial\over\partial r^{\mu}}\int_{0}^{t}d\tau\int d^{2}{\bf
r}_{1}P_{eq}({\bf r}) { V}^{\mu}(1\rightarrow 0,t){ V}^{\nu}(1\rightarrow 0,
t-\tau)\nonumber\\
 \times\biggl ({\partial P \over\partial r^{\nu}}({\bf
r}(t-\tau),t-\tau)+\beta_{eq}\gamma P({\bf r}(t-\tau),t-\tau)   {\partial
\psi_{eq}\over\partial r^{\nu}}({\bf r}(t-\tau))\biggr ).
\label{K5}
\end{eqnarray}

\subsection{The Fokker-Planck equation}
\label{sec_FPeq}

\subsubsection{Unidirectionnal flow}
\label{sec_uniuni}

We will now see how the previous equation can be simplified for particular
equilibrium flows. We shall first consider the case of an unidirectional flow
$\langle {\bf V}\rangle_{eq}=V_{eq}(y)\hat {\bf x}$ produced by a vorticity
distribution $\langle \omega\rangle_{eq}(y)$. If we restrict ourselves to
solutions of the form $P=P(y,t)$, the kinetic equation (\ref{K5}) becomes 
\begin{eqnarray}
{\partial P\over \partial t}= N {\partial\over\partial y}\int_{0}^{t}d\tau\int
d^{2}{\bf r}_{1} P_{eq}(y){ V}^{y}(1\rightarrow 0,t){ V}^{y}(1\rightarrow 0,
t-\tau)\nonumber\\
 \times\biggl ({\partial P \over\partial
y}(y,t-\tau)+\beta_{eq}\gamma  P(y,t-\tau)  {\partial \psi_{eq}\over\partial
y}({y})\biggr ),
\label{uniK1}
\end{eqnarray}  
where we have used $y(t-\tau)=y(t)=y$ since the point vortices follow the 
streamlines of the equilibrium flow. This equation can be rewritten in the form
\begin{eqnarray}
{\partial P\over \partial t}={\partial\over\partial y}\int_{0}^{t} d\tau C(\tau)
\biggl ( {\partial P \over\partial y}(y,t-\tau)+\beta_{eq}\gamma P(y,t-\tau)
{\partial\psi_{eq}\over\partial y}\biggr ),
\label{uninonmark}
\end{eqnarray}  
where
\begin{eqnarray}
C(\tau)\equiv C^{yy}(\tau)=N\int d^{2}{\bf r}_{1}{ V}^{y}(1\rightarrow 0,t) {
V}^{y}(1\rightarrow 0, t-\tau)P_{eq}(y)
\label{yto}
\end{eqnarray}  
is the velocity autocorrelation function. In Appendix \ref{sec_uni}, it is found
that:
\begin{eqnarray}
C(\tau)={N\gamma^{2}\over 8\pi}\ln N {1\over 1+{1\over 4}\Sigma^{2}(y)\tau^{2}}
P_{eq}(y),
\label{uni9}
\end{eqnarray} 
where 
\begin{eqnarray}
\Sigma(y)= -{d\over dy} \langle V\rangle_{eq}(y)
\label{uni6}
\end{eqnarray} 
is the local shear of the flow, equal here to the vorticity. Note that the
velocity autocorrelation function decays like $\tau^{-2}$ for $\tau\rightarrow
+\infty$ \cite{chav98}. 

Equation (\ref{uninonmark}) is a non Markovian equation since the probability
$P(y,t)$ at time $t$ depends on the probability $P(y,t-\tau)$ at earlier times
through an integration over $\tau$. Accordingly, the present study which
explicitly takes into account memory effects is more general than the stochastic
model presented in section \ref{sec_fp}. However, if we implement a Markov
approximation and replace $P(y,t-\tau)$ by $P(y,t)$, we recover the
Fokker-Planck equation of section \ref{sec_elementary}:
\begin{eqnarray}
{\partial P\over \partial t}={\partial\over\partial y}\biggl \lbrack D\biggl (
{\partial P\over\partial y}+\beta_{eq}\gamma P {\partial\psi_{eq}\over\partial
y}\biggr )\biggr \rbrack 
\label{uniFP}
\end{eqnarray} 
with a diffusion coefficient
\begin{eqnarray}
D\equiv D^{yy}=N\int_{0}^{+\infty}d\tau\int d^{2}{\bf r}_{1}{
V}^{y}(1\rightarrow 0,t) { V}^{y}(1\rightarrow 0, t-\tau)P_{eq}(y)
\label{uniD}
\end{eqnarray} 
and a drift term 
\begin{eqnarray}
\eta_{y}\equiv -\langle V_{y}\rangle_{drift}= \beta_{eq}\gamma D {\partial
\psi_{eq}\over\partial y}.
\label{unieta}
\end{eqnarray} 
The drift coefficient is given by an Einstein relation as expected from the
general considerations of section \ref{sec_fp}. The diffusion coefficient is
expressed as a Kubo formula, i.e. as the integral of the velocity correlation
function (see Appendix \ref{sec_diffcoeff}). In Appendix \ref{sec_uni} it is
found that (see also Ref. \cite{chav98}):
\begin{eqnarray}
D={1\over 8}N\gamma^{2}{1\over |\Sigma(y)|}\ln N P_{eq}(y).
\label{uni11}
\end{eqnarray} 
The reason for the logarithmic divergence is explained in Appendix
\ref{sec_statistics} and in Ref. \cite{cs}.

\subsubsection{Axisymmetrical flow}
\label{sec_axiaxi}

We now consider the case of an axisymmetrical equilibrium flow such that
$\langle {\bf V}\rangle_{eq}=\langle {V}\rangle_{eq}(r){\hat{\bf e}}_{\theta}$.
This flow is generated by an equilibrium vorticity field
$\langle\omega\rangle_{eq}(r)$. If we restrict ourselves to solutions of the
form $P=P(r,t)$, the kinetic equation (\ref{K5}) simplifies in  
\begin{eqnarray}
{\partial P\over \partial t}= N{1\over r} {\partial\over\partial
r}r\int_{0}^{t}d\tau\int d^{2}{\bf r}_{1}P_{eq}(r){ V}^{r(t)}(1\rightarrow 0,t){
V}^{r(t-\tau)}(1\rightarrow 0, t-\tau)\nonumber\\ 
\times\biggl ({\partial P
\over\partial r}(r,t-\tau)+\beta_{eq}\gamma  P(r,t-\tau)  {\partial
\psi_{eq}\over\partial r}({r})\biggr ),
\label{Kaxi1}
\end{eqnarray}  
where we have used $r(t-\tau)=r(t)=r$ since the point vortices follow the
streamlines of the equilibrium flow. Furthermore, $V^{r(t)}$ and $V^{r(t-\tau)}$
denote the radial components of ${\bf V}(1\rightarrow 0)$ at times $t$ and
$t-\tau$ in a polar system of coordinates (${\hat{\bf e}}_{r}(t),{\hat{\bf
e}}_{\theta}(t)$) moving with the test vortex. Equation (\ref{Kaxi1}) can be
rewritten 
\begin{eqnarray}
{\partial P\over \partial t}={1\over r}{\partial\over\partial r}
r\int_{0}^{t}d\tau C(\tau) \biggl ( {\partial P \over\partial
r}(r,t-\tau)+\beta_{eq}\gamma P(r,t-\tau) {\partial\psi_{eq}\over\partial
r}\biggr ),
\label{Kaxi5}
\end{eqnarray} 
where $ C(\tau)$ is the velocity autocorrelation function
\begin{eqnarray}
C(\tau)=N\int d^{2}{\bf r}_{1}{ V}^{r(t)}(1\rightarrow 0,t) {
V}^{r(t-\tau)}(1\rightarrow 0, t-\tau)P_{eq}(r).
\label{fgr}
\end{eqnarray} 
In Appendix \ref{sec_axi} it is found that
\begin{eqnarray}
C(\tau)={N\gamma^{2}\over 8\pi}\ln N {1\over 1+{1\over 4}\Sigma^{2}(r)\tau^{2}}
P_{eq}(r),
\label{axi11}
\end{eqnarray}
where 
\begin{equation}
\Sigma(r)=r {d\over dr}\biggl({\langle V\rangle_{eq}(r)\over r}\biggr )
\label{axi9}
\end{equation}
is the local shear of the flow.

If we ignore memory effects, we obtain the Fokker-Planck equation: 
\begin{eqnarray}
{\partial P\over \partial t}={1\over r}{\partial\over\partial r}\biggl \lbrack r
D\biggl ( {\partial P\over\partial r}+\beta_{eq}\gamma P
{\partial\psi_{eq}\over\partial r}\biggr )\biggr \rbrack 
\label{Kaxi4}
\end{eqnarray} 
with a diffusion coefficient 
\begin{eqnarray}
D=N\int_{0}^{+\infty}d\tau\int d^{2}{\bf r}_{1}{ V}^{r(t)}(1\rightarrow 0,t)
{ V}^{r(t-\tau)}(1\rightarrow 0, t-\tau)P_{eq}(r)
\label{Kaxi2}
\end{eqnarray} 
and a drift term
\begin{eqnarray}
\eta_{r}=-\langle V_{r}\rangle_{drift}=\beta_{eq}\gamma D
{\partial\psi_{eq}\over\partial r}.
\label{Kaxi3}
\end{eqnarray} 
Explicitly, the diffusion coefficient has the form (see Appendix \ref{sec_axi}
and Ref. \cite{chav98}):
\begin{eqnarray}
D={1\over 8}N\gamma^{2}{1\over |\Sigma(r)|}\ln N P_{eq}(r).
\label{axi12}
\end{eqnarray}

\subsubsection{The general case}
\label{sec_summary}

We can show in the general case that the relaxation of the test vortex is 
described by the Fokker-Planck equation 
\begin{eqnarray}
{\partial P\over\partial t}+\langle {\bf V}\rangle_{eq} \nabla P=\nabla(D(\nabla
P+\beta_{eq}\gamma P\nabla\psi_{eq}))
\label{wr1}
\end{eqnarray} 
with a diffusion coefficient
\begin{eqnarray}
D={\gamma\over 8}{1\over |\Sigma({\bf r})|}\ln N\langle\omega\rangle_{eq},
\label{wr2}
\end{eqnarray} 
where $|\Sigma({\bf r})|=2\sqrt{-Det(\Sigma)}$ is the local shear of the flow
and and $Det(\Sigma)$ the determinant of the stress tensor $\Sigma^{\mu\nu}$
(see section \ref{sec_connexionFP}).  In the regions where the shear cancels
out, our approximations clearly break up. In particular, we cannot calculate the
Kubo integral by assuming that the vortices follow the streamlines of the
equilibrium flow. This is because, for a local solid rotation, the vortices
always remain at the same relative distance and the correlation time is
infinite. In that case it is necessary to take into account the dispersion of
the vortices. An alternative derivation of the diffusion coefficient  can be
obtained by analyzing the statistics of velocity fluctuations created by a
random distribution of point vortices \cite{cs}. When the differential rotation
of the vortices is neglected (which corresponds to the opposite limit of that
leading to equation (\ref{wr2})), we obtain (see Ref. \cite{cs} and Appendix
\ref{sec_statistics}): 
\begin{eqnarray}
D\sim \gamma \sqrt{\ln N}  
\label{conc5} 
\end{eqnarray} 
Clearly, a more complete study should take into account simultaneously the
effect of the shear and the dispersion of the vortices to match the two formulae
(\ref{wr2}) and (\ref{conc5}).

For $t\rightarrow +\infty$, the distribution function $P({\bf r},t)$ of the test
vortex converges towards the Maxwell-Boltzmann statistics (\ref{jm}). The time
of relaxation corresponds typically to the time needed by the test vortex to
diffuse over a distance $R$, the system size. Therefore $t_{relax}\sim R^{2}/D$
where $D\sim \gamma\ln N$ is the order of magnitude of the diffusion coefficient
given by equation  (\ref{wr2}). Using $\Gamma=N\gamma$ and introducing the
dynamical time $t_{D}\sim\langle\omega\rangle^{-1}\sim R^{2}/\Gamma$, we obtain
the estimate:
\begin{eqnarray}
t_{relax}\sim {{ N}\over\ln { N}}t_{D}.
\label{conc6}
\end{eqnarray} 
Since the statistical description is expected to yield relevant results for
large $N$, we conclude  that the relaxation of point vortices towards the
Boltzmann distribution is a very slow process. It is plausible that a more
violent relaxation be at work in the system. This problem is discussed more
specifically in the conclusion.

In the previous calculations, we have assumed that the distribution
of the field vortices is given by the Maxwell-Boltzmann statistics (\ref{jm})
which corresponds to statistical equilibrium. In the case of an arbitrary background distribution $P_{eq}$, the expression of the  drift
is
\begin{eqnarray}
\langle {\bf V}\rangle_{drift}=D\nabla \ln P_{eq},
\label{aj1}
\end{eqnarray} 
where $D$ is still given by equation (\ref{wr2}). Since $D>0$, the drift is
always directed along the density gradient. The estimate of the time of
relaxation is not changed in this more general situation.

\section{A generalized kinetic equation}
\label{sec_generalized}

In the previous section, we have described the relaxation of a test vortex in a
``bath'' of field vortices at statistical equilibrium. We would like now to
relax this thermal bath approximation in order to obtain a generalized kinetic
equation describing the evolution of the whole system.

\subsection{The factorization hypothesis}
\label{sec_factorization}

If the vortices are initially decorrelated then, for sufficiently short times,
they will remain decorrelated. This means that the $(N+1)$-particle distribution
function can be factorized in a product of ($N+1$) one-particle distribution
functions:
\begin{equation}
\mu({\bf r},{\bf r}_{1},...,{\bf r}_{N},t)=\prod_{k=0}^{N} P({\bf r}_{k},t).
\label{decorr}
\end{equation}
If we integrate the Liouville equation (\ref{Liouville1}) on the positions of
the $N$ vortices $1,...,N$ and use the factorization (\ref{decorr}), we directly
obtain 
\begin{equation}
{\partial P\over\partial t}+\langle {\bf V}\rangle\nabla P=0.
\label{Eulershort}
\end{equation} 
Therefore, for sufficiently short times, the average vorticity
$\langle\omega\rangle$ satisfies the 2D Euler equation. However, at later times,
the distribution function $\mu$ differs from the pure product (\ref{decorr}) and
the Euler equation does not provide a good approximation anymore.  In section
\ref{sec_liouville} we have determined an exact equation (\ref{kin2}) satisfied
by the one-particle distribution function at any time. This equation is not
closed, however, since it involves the $N$-vortex distribution function
$\mu_{sys}$. We shall close the system by assuming that $\mu_{sys}$ can still be
approximated by  a product of N one-particle distribution functions:
\begin{equation}
\mu_{sys}({\bf r}_{1},...,{\bf r}_{N},t)\simeq \prod_{k=1}^{N} P({\bf r}_{k},t)
\label{decorrsys}
\end{equation}  
but, contrary to section \ref{sec_bath}, the one-particle distribution function
$P({\bf r}_{k},t)$ is {\it not} ascribed to the equilibrium value $P_{eq}$.
Physically, the decomposition (\ref{muI}) (\ref{decorrsys}) assumes that the
correlations that develop between the vortices (term $\mu_{I}$) are due uniquely
to the polarization cloud imposed by each individual vortex. Without this
polarization, the vortices would be uncorrelated  (term $\mu_{sys}$). In
particular, this decomposition does not take into account three-body encounters
which can play a crucial role in the dynamics of vortices (in particular 
when the system is neutral and homogeneous, see Ref. \cite{novikov,sc}). These 
high order correlations develop on longer time scales and may be neglected in 
a first approach.

The approximation (\ref{decorrsys}) introduced in equation (\ref{kin2}) leads to
a generalized kinetic equation  
\begin{eqnarray}
{\partial P\over\partial t}+\langle {\bf V}\rangle{\partial P\over\partial {\bf
r}}={\partial\over\partial r^{\mu}}\int\prod_{k=1}^{N}d^{2}{\bf
r}_{k}\int_{0}^{t}d\tau\sum_{i=1}^{N}\sum_{j=1}^{N}  {\cal V}^{\mu}(i\rightarrow
0)\nonumber\\ 
\times{\cal G}(t,t-\tau)\biggl ({\cal V}^{\nu}(j\rightarrow
0){\partial\over\partial {r}^{\nu}}+{{\cal V}}^{\nu}(0\rightarrow
j){\partial\over\partial { r}_{j}^{\nu}}\biggr ) P({\bf
r},t-\tau)\prod_{k=1}^{N} P({\bf r}_{k},t-\tau).
\label{genK1}
\end{eqnarray}
Repeating the steps leading from equation (\ref{K1}) to equation (\ref{K4}), we
obtain 
\begin{eqnarray}
{\partial P\over\partial t}+\langle {\bf V}\rangle{\partial P\over\partial {\bf
r}}=N{\partial\over\partial r^{\mu}}\int_{0}^{t}d\tau\int d^{2}{\bf
r}_{1}{V}^{\mu}(1\rightarrow 0)_{t}\nonumber\\
 \times\biggl \lbrace
{V}^{\nu}(1\rightarrow 0)P_{1}{\partial P\over\partial {r}^{\nu}}+{{
V}}^{\nu}(0\rightarrow 1)P{\partial P_{1}\over\partial { r}_{1}^{\nu}}\biggr
\rbrace_{t-\tau}, 
\label{new1}
\end{eqnarray}
where $P=P({\bf r},t)$ and $P_{1}=P({\bf r}_{1},t)$. We also recall that between
$t$ and $t-\tau$ the trajectories of the particles are determined from the
smooth velocity field created by the vorticity distribution
$\langle\omega\rangle=N\gamma P({\bf r},t)$. This non Markovian
integrodifferential equation is similar to the equation obtained by Chavanis
(2000) \cite{chavquasi} in two-dimensional turbulence using a quasilinear theory
of the Euler-Poisson system. It is shown in Appendix \ref{sec_conservation} that
this equation rigorously conserves angular momentum in a circular domain and
linear impulse in a channel (or in an infinite domain). However, under this
form, it is not possible to prove the conservation of energy and the H-theorem.
In our previous investigation of the problem \cite{chavquasi}, we have
considered a somewhat crude approximation which amounts to neglecting memory
effects in equation (\ref{new1}). We beleived that this approximation would not
introduce any significant error but, as we shall see, we were wrong: such
approximation breaks the conservation of energy.

If we assume that the decorrelation time $\tau$ is short (which does not need to
be the case) and implement a strong Markov approximation, we obtain 
\begin{eqnarray}
{\partial P\over\partial t}+\langle {\bf V}\rangle{\partial P\over\partial {\bf
r}}={N\tau\over 2}{\partial\over\partial r^{\mu}}\int d^{2}{\bf
r}_{1}{V}^{\mu}(1\rightarrow 0) \biggl \lbrace {V}^{\nu}(1\rightarrow
0)P_{1}{\partial P\over\partial {r}^{\nu}}+{{ V}}^{\nu}(0\rightarrow
1)P{\partial P_{1}\over\partial { r}_{1}^{\nu}}\biggr \rbrace.
\label{new2}
\end{eqnarray}
In the case of an infinite domain  ${\bf V}(0\rightarrow 1)=-{\bf
V}(0\rightarrow 1)$ and we have the further simplification 
\begin{eqnarray}
{\partial P\over\partial t}+\langle {\bf V}\rangle{\partial P\over\partial {\bf
r}}={N\gamma^{2}\over 8\pi^{2}}\tau{\partial\over\partial r^{\mu}}\int d^{2}{\bf
r}_{1} K'^{\mu\nu}({\mb\xi})\biggl ( P_{1}{\partial P\over\partial {r}^{\nu}}-
P{\partial P_{1}\over\partial { r}_{1}^{\nu}}\biggr ),
\label{genK4}
\end{eqnarray}
where 
\begin{eqnarray}
K'^{\mu\nu}({\mb\xi})={\xi_{\perp}^{\mu}\xi_{\perp}^{\nu}\over\xi^{4}}=
{\xi^{2}\delta^{\mu\nu}-\xi^{\mu}\xi^{\nu}\over\xi^{4}}
\label{Kmunu}
\end{eqnarray}
and ${\mb\xi}={\bf r}_{1}-{\bf r}$ (we also recall that ${\mb\xi}_{\perp}$ is
the vector ${\mb\xi}$ rotated by $+{\pi\over 2}$). To arrive at equation
(\ref{Kmunu}) we have explicitly  used the form of the Kernel (\ref{Vji}) and to
get the second equality we have used the fact that we are in two dimensions.
Note that the symmetrical form of equation (\ref{genK4}) is reminiscent of the
Landau equation introduced in plasma physics and in stellar dynamics
\cite{kandrup}. In this analogy, the position ${\bf r}$ of the vortices plays
the role of the velocity ${\bf v}$ of the electric charges or stars. Therefore,
we can directly infer the conservation  of linear impulse ${\bf
P}_{\perp}=\int\omega {\bf r} d^{2}{\bf r}$ and angular momentum $L=\int \omega
r^{2}d^{2}{\bf r}$ which play respectively the role of impulse ${\bf P}=\int f
{\bf v} d^{3}{\bf v}$ and kinetic energy $K=\int f {v^{2}\over 2}d^{3}{\bf v}$
in plasma physics. In addition, we can prove a H-theorem for the Boltzmann
entropy (\ref{entropy}) exactly like for the Landau equation. Finally, we can
show that the solutions of equation (\ref{genK4}) converge towards the  Gaussian
vortex (the equivalent of the Maxwellian distribution in plasma physics with
${\bf r}$ in place of ${\bf v}$):
\begin{eqnarray}
P({\bf r})=A e^{-{1\over 2}\alpha\gamma ({\bf r}-{\bf r}_{0})^{2}}
\label{Gauss}
\end{eqnarray}
which is the maximum entropy state at fixed circulation, angular momentum and
impulse. It is in general different from the Boltzmann distribution (\ref{jm})
with the relative streamfunction $\psi'=\psi+{\Omega\over 2}r^{2}-{\bf
U}_{\perp}.{\bf r}$. This clearly indicates that equation (\ref{new2}) does {\it
not} conserve the meanfield  energy (\ref{E2}). It may happen, however, that the
energy is approximately conserved if we start from an initial condition with a
value of energy $E_{0}$ corresponding to $\beta\rightarrow 0$,
$\Omega\rightarrow +\infty$ and $\alpha=\beta\Omega/2$ finite at equilibrium. In
that case, equation (\ref{Gauss}) is the maximum entropy state at fixed
$\Gamma$, $L$ and $E_{0}$ (the conservation of impulse can be satisfied
trivially by taking the center of vorticity as origin of our system of
coordinates). However, if $E$  differs from $E_{0}$ by a large amount, the
kinetic equation (\ref{new2}) will not correctly describe the evolution of the
system for late times.

Now, if we account properly for memory effects in equation (\ref{new1}), we
can obtain a more general equation which guaranties in addition the conservation
of energy and is therefore more satisfactory. In the case of an axisymmetrical
flow, it is possible to calculate the memory function appearing in equation
(\ref{new1}) explicitly if we assume that the correlation time is smaller (but
not necessarily {\it much} smaller) than the typical time on which the average
vorticity changes appreciably. In this approximation, the point vortices follow,
between $t$ and $t-\tau$, circular trajectories with angular velocity
$\Omega(r,t)=\langle V_{\theta}\rangle(r,t)/r$ and equation (\ref{new1})
simplifies in (see Appendix \ref{sec_memory}):
\begin{eqnarray}
{\partial P\over\partial t}=-{N\gamma^{2}\over 4 r}{\partial\over\partial
r}\int_{0}^{+\infty}r_{1}dr_{1} \delta(\Omega-\Omega_{1})\ln\biggl \lbrack
1-\biggl ({r_{<}\over r_{>}}\biggr )^{2}\biggr \rbrack\biggl\lbrace {1\over
r}P_{1}{\partial P\over\partial r}- {1\over r_{1}}P{\partial P_{1}\over\partial
r_{1}} \biggr\rbrace,   
\label{day1}
\end{eqnarray}
where $\Omega=\Omega(r,t)$, $\Omega_{1}=\Omega(r_{1},t)$ and $r_{>}$ (resp.
$r_{<}$) is the biggest (resp. smallest) of $r$ and $r_{1}$. The angular velocity is related to the vorticity by
\begin{eqnarray}
\langle\omega\rangle ={1\over r}{\partial\over\partial r}(\Omega r^{2}).
\label{qw1}
\end{eqnarray}
Similarly, in the case of a unidirectional flow, we can assume that, between $t$
and $t-\tau$,  the point vortices follow linear trajectories with velocity
$\langle {\bf V}\rangle=\langle V\rangle(y,t) {\bf e}_{x}$. This leads to the
kinetic equation  (see Appendix \ref{sec_memory}):
\begin{eqnarray}
{\partial P\over\partial t}={N\gamma^{2}\over 4}{\partial\over\partial
y}\int_{-\infty}^{+\infty}dy_{1}\delta (V_{1}- V)E_{1}\biggl ({2|y_{1}-y|\over
L}\biggr )\biggl (P_{1}{\partial P\over\partial y}-P{\partial P_{1}\over\partial
y_{1}}\biggr ),
\label{day5}
\end{eqnarray}
where $V =\langle V\rangle (y,t)$ and $V_{1} =\langle V\rangle (y_{1},t)$. The
function $E_{1}(x)$ is the exponential integral and $L$ an upper cut-off
necessary in that case (see Appendix \ref{sec_memory}). We also recall that the average velocity is related to the vorticity by
\begin{eqnarray}
\langle\omega\rangle =-{\partial\over\partial y}\langle V\rangle.
\label{qw2}
\end{eqnarray}

We can remarkably propose an approximation of the general kinetic equation (\ref{new1}) which encompasses both the axisymmetric form (\ref{day1}) and the unidirectional form (\ref{day5}). Memory effects are not neglected, unlike in Eq. 
(\ref{genK4}), but they are simplified in a way which preserves all the conservation laws of the system (as discussed below). We propose the generalized kinetic equation:
\begin{eqnarray}
{\partial P\over\partial t}+\langle {\bf V}\rangle \nabla P={N\gamma^{2}\over
8}{\partial \over\partial r^{\mu}}\int d^{2}{\bf r}_{1}K^{\mu\nu}\delta ({\mb
\xi}.{\bf v})\biggl (P_{1}{\partial P\over\partial r^{\nu}}-P{\partial
P_{1}\over\partial r_{1}^{\nu}}\biggr )
\label{sun1}
\end{eqnarray}
with  
\begin{eqnarray}
K^{\mu\nu}({\mb\xi})={\xi_{\perp}^{\mu}\xi_{\perp}^{\nu}
\over\xi^{2}}={\xi^{2}\delta^{\mu\nu}-\xi^{\mu}\xi^{\nu}\over \xi^{2}}
\label{Kmunu2}
\end{eqnarray}
and ${\mb\xi}={\bf r}_{1}-{\bf r}$, ${\bf v}=\langle {\bf V}\rangle ({\bf
r}_{1},t)-\langle {\bf V}\rangle ({\bf r},t)$.  For specific applications, it may
be necessary to introduce a shielding of the form (\ref{u16}) in the interaction
between vortices. This shielding arises naturally in geophysics in the
``quasigeostrophic approximation''. In that case the tensor $K^{\mu\nu}$ is replaced
by $(1/L)K_{1}(\xi/L)\xi_{\perp}^{\mu}\xi_{\perp}^{\nu}/\xi$ where 
$K_{1}$ is the modified Bessel function of first order and  $L$ is an
upper-cut off (called the Rossby radius in geophysics) which plays 
the same role
as the Debye length in plasma  physics.

Equation (\ref{sun1}) is not exact, in a strict sense, yet it satisfies all the
conservations laws of the vortex system as well as a H-theorem (section
\ref{sec_lawsH}). This is very gratifying and this can have important practical
applications. It is remarkable that we can write down an approximate kinetic
equation in the general case without being required to specify the trajectory of
the point vortices between $t$ and $t-\tau$. In fact, to arrive at equation
(\ref{sun1}) we have made implicitly two approximations: (i) we have assumed 
that the
vorticity field does not change dramatically when we follow the vortices in their
motion between $t$ and $t-\tau$. (ii) After the time integration has been effected, the non
universal function $V^{\nu}(1\rightarrow 0,t-\tau)$ gives rise to a logarithmic
term which has been replaced by $1$ in the subsequent calculations. This term
produces a sub-logarithmic correction that is
flow-dependant and that has been neglected. It is on account of 
this weak dependance that a general kinetic
equation can be obtained. For axisymmetrical or unidirectionnal flows, equation
(\ref{sun1}) reduces to equations (\ref{day1}) and (\ref{day5}) with
$\ln(|r_{1}-r|/(r_{1}+r))$ and $K_{0}(|y-y_{1}|/L)$ instead of
$\ln(1-(r_{<}/r_{>})^{2})$ and $E_{1}(2|y-y_{1}|/L)$. This is not a too severe
discrepency  (these functions have a similar logarithmic behaviour) so our
approximations are reasonable.

From equations (\ref{day1}) (\ref{day5}) and (\ref{sun1}), it is clear that the
relaxation towards equilibrium is due to a phenomenon of resonance. Only the
points ${\bf r}_{1}$ satisfying the condition ${\mb \xi}.{\bf v}=0$ 
with ${\bf r}_{1}\neq {\bf r}$ will
contribute to the diffusion current in ${\bf r}$. In the axisymmetrical case,
this condition of resonance reduces to $\Omega(r_{1})=\Omega(r)$ and in the
unidirectional case to $V(y_{1})=V(y)$. These conditions of resonance had never
been noticed previously. Further work on the subject will have to make these
criteria more precise by computing explicitly ``resonance lines'' in
two-dimensional real flows.

\subsection{Conservation laws and H-theorem}
\label{sec_lawsH}

We now derive the conservation laws and the H-theorem satisfied by equation
(\ref{sun1}). First of all, the conservation of the circulation is
straightforward since equation (\ref{sun1}) can be written in the form of a
continuity equation (\ref{c1}). To prove the conservation of angular momentum,
we start from equation (\ref{c4}), substitute for (\ref{sun1}), permut the dummy
variables ${\bf r}$ and ${\bf r}_{1}$ and add the resulting expressions. This
yields
\begin{eqnarray}
\dot L={N^{2}\gamma^{3}\over 8}\int d^{2}{\bf r}d^{2}{\bf
r}_{1}K^{\mu\nu}\xi^{\mu}\delta ({\mb\xi}.{\bf v})\biggl  (P_{1}{\partial
P\over\partial r^{\nu}}-P{\partial P_{1}\over\partial r_{1}^{\nu}}\biggr ).
\label{mon3}
\end{eqnarray}
But, from equation (\ref{Kmunu2}), we immediately verify that
\begin{eqnarray}
K^{\mu\nu}\xi^{\mu}=0.
\label{mont3}
\end{eqnarray}
This proves the conservation of angular momentum. We can prove the conservation
of linear impulse in a similar manner. Starting from equation (\ref{c16}),
substituting for (\ref{sun1}), permuting the dummy variables ${\bf r}$ and ${\bf
r}_{1}$ and adding the two resulting expressions yields $\dot {\bf P}={\bf 0}$.
For the conservation of energy, we start from equation (\ref{dotE}) and follow
the by-now familiar procedure. This yields
\begin{eqnarray}
\dot E={N^{2}\gamma^{3}\over 16}\int d^{2}{\bf r}d^{2}{\bf
r}_{1}K^{\mu\nu}v_{\perp}^{\mu}\delta ({\mb\xi}.{\bf v})\biggl  (P_{1}{\partial
P\over\partial r^{\nu}}-P{\partial P_{1}\over\partial r_{1}^{\nu}}\biggr ).
\label{mon7}
\end{eqnarray}
Considering the form of the tensor (\ref{Kmunu2}), we have 
\begin{eqnarray}
K^{\mu\nu}v_{\perp}^{\mu}={\xi_{\perp}^{\nu}\over \xi^{2}}({\mb\xi}.{\bf v}).
\label{mon8}
\end{eqnarray}
When substituted in equation (\ref{mon7}), we see that the occurence of the
delta function in the kinetic equation  implies $\dot E=0$. Finally, for the
rate of entropy production we have, according to equations (\ref{dotS}) and
(\ref{sun1}):
\begin{eqnarray}
\dot S={N^{2}\gamma^{2}\over 8}\int d^{2}{\bf r}d^{2}{\bf r}_{1}{1\over P
P_{1}}P_{1}{\partial P\over\partial r^{\mu}} K^{\mu\nu}\delta ({\mb\xi}.{\bf
v})\biggl  (P_{1}{\partial P\over\partial r^{\nu}}-P{\partial P_{1}\over\partial
r_{1}^{\nu}}\biggr ).
\label{mon9}
\end{eqnarray}
Permutting the dummy variables ${\bf r}$ and ${\bf r}_{1}$ and adding the
resulting expression to equation (\ref{mon9}), we get
\begin{eqnarray}
\dot S={N^{2}\gamma^{2}\over 16}\int d^{2}{\bf r}d^{2}{\bf r}_{1}{1\over P
P_{1}}\delta ({\mb\xi}.{\bf v}) \biggl  (P_{1}{\partial P\over\partial
r^{\mu}}-P{\partial P_{1}\over\partial r_{1}^{\mu}}\biggr )  K^{\mu\nu}\biggl
(P_{1}{\partial P\over\partial r^{\nu}}-P{\partial P_{1}\over\partial
r_{1}^{\nu}}\biggr ).
\label{mon10}
\end{eqnarray}
Now, for any vector, $A^{\mu}K^{\mu\nu}A^{\nu}=({\bf
A}{\mb\xi}_{\perp})^{2}/\xi^{2}\ge 0$. This proves a H-theorem ($\dot S\ge 0$)
for our kinetic equation (\ref{sun1}). It should be emphasized that the
conservation laws and the H-theorem result from the {\it symmetry} of the
kinetic equation (and the condition of resonance) and not from formal  Lagrange
multipliers like in the thermodynamical approach of section \ref{sec_mepp}. In
addition the H-theorem is {\it proved} by our approach instead of beeing
postulated.  This is more satisfying on physical grounds. 

It remains for one to show that the Boltzmann distribution
\begin{eqnarray}
P=Ae^{-\beta\gamma (\psi+{\Omega\over 2}r^{2}-{\bf U}_{\perp}.{\bf r})}
\label{mon11}
\end{eqnarray}
is a stationary solution of equation (\ref{sun1}). Noting that
\begin{eqnarray}
{\partial P\over\partial r^{\nu}}=-\beta\gamma\biggl ({\partial\psi\over\partial
r^{\nu}}+\Omega r^{\nu}-U_{\perp}^{\nu}\biggr )P,
\label{mon12}
\end{eqnarray}
we have successively
\begin{eqnarray}
K^{\mu\nu}\biggl (P_{1}{\partial P\over\partial r^{\nu}}-P{\partial
P_{1}\over\partial r_{1}^{\nu}}\biggr )=\beta\gamma
PP_{1}K^{\mu\nu}(v_{\perp}^{\nu}+\Omega\xi^{\nu})=\beta\gamma
PP_{1}{\xi_{\perp}^{\mu}\over \xi^{2}}({\mb\xi}.{\bf v}),
\label{mon13}
\end{eqnarray}
where we have used equations (\ref{mont3}) and (\ref{mon8}). When  substituted
in equation (\ref{sun1}), we find that the r.h.s. cancels out due to the delta
function. The advective term is also zero since $P=f(\psi')$. Therefore, the
distribution (\ref{mon11}) is a stationary solution of equation (\ref{sun1}).
Note, however, that this is not the only solution. Any stationary solution
satisfying ${\mb\xi}.{\bf v}\neq 0$ for any couple of points ${\bf r}$, ${\bf
r}_{1}$ (with  ${\bf r}\neq {\bf r}_{1}$) is a solution of equation
(\ref{sun1}). Physically, this implies that the system needs sufficiently strong
resonances to relax towards the maximum entropy state. If this is not realized
it will be frozen in a sort of ``metastable'' equilibrium. This may explain why
the maximum entropy state is not always reached in two-dimensional turbulence.
For example, for a unidirectional flow with $\langle\omega\rangle$ positive or
negative everywhere, the velocity field is monotonous  (see equation (\ref{qw2}))
and the condition of resonance cannot be satisfied. The evolution of the
system will require non trivial correlations between  point vortices that
are not taken into account in the present approch. One would need to replace the
factorization hypothesis (\ref{decorrsys}) by a product of two-point or
three-point correlations functions. However, it is plausible that these
correlations develop on a very long time scale so it remains a  matter of debate
to decide whether they really are relevant for the dynamics or not.

In the context of 2D turbulence described by the Euler-Poisson system, the
quasilinear theory developed by Chavanis (2000) yields instead of equation
(\ref{sun1}):
\begin{eqnarray}
{\partial \overline{\omega}\over\partial t}+{\bf
u}\nabla\overline{\omega}={\epsilon^{2}\over 8}{\partial\over\partial
r^{\mu}}\int d^{2}{\bf r}'K^{\mu\nu}\delta({\mb\xi}.{\bf v})\biggl\lbrace
\overline{\omega}'(\sigma_{0}-\overline{\omega}')
{\partial\overline{\omega}\over\partial
r^{\mu}}-\overline{\omega}(\sigma_{0}-\overline{\omega})
{\partial\overline{\omega}'\over\partial r'^{\mu}}\biggr\rbrace,
\label{mon14}
\end{eqnarray}
where $K^{\mu\nu}$ is defined by (\ref{Kmunu2}) and
$\overline{\omega}=\overline{\omega}({\bf r},t)$,
$\overline{\omega}'=\overline{\omega}({\bf r}',t)$. In addition to the previous
conservation laws, this equation guaranties that the coarse-grained vorticity
$\overline{\omega}$ remains bounded by the maximum value of the initial
distribution, i.e. $\overline{\omega}\le\sigma_{0}$. This equation satisfies a
H-theorem for the Fermi-Dirac entropy introduced by Miller-Robert-Sommeria at
equilibrium   \cite{miller,rseq}. Our approach provides therefore another way of
justifying their results from a dynamical point of view. Equation (\ref{mon14})
is written for a single level of vorticity $\sigma_{0}$, but it is possible to
extend the quasilinear theory to an arbitrary distribution of vorticity levels
(in preparation). Our equations should provide therefore an interesting and
useful parametrization of the 2D Euler equation. It should be recalled in that
respect that the usual turbulent diffusion $\nu\Delta\overline{\omega}$
introduced {\it ad hoc} in the r.h.s. of the 2D Euler equation in order to
smooth out the small scales and prevent numerical instabilities breaks the
conservation laws of the inviscid dynamics. This is {\it not} the case for our
equation (\ref{mon14}): not only it smoothes out the unresolved scales (as
exemplified by the existence of a H-theorem) but it satisfies all the
conservation laws of the inviscid dynamics and respects the invariance
properties of the Euler equation (invariance by translation and rotation of the
coordinates, Galilean invariance and invariance by rotation of the referential).
In addition there is no free parameter in our theory except the coarse-graining
mesh $\epsilon$ (or resolution scale) which depends on the situation
contemplated. Different attempts had been made previously to obtain an equation
satisfying all these requirements, but only partial results were obtained
\cite{rs,csr,cs97,chav98,chavquasi}.

\subsection{Connexion with the Fokker-Planck equation}
\label{sec_connexionFP}

Equation (\ref{sun1}) can be considered as our final result but we wish to show
that a direct connection with the Fokker-Planck equation of section
\ref{sec_relaxation} can be found.  Introducing a diffusion tensor
\begin{equation}
D^{\mu\nu}={N\gamma^{2}\over 8} \int d^{2}{\bf r}_{1}
K^{\mu\nu}\delta({\mb\xi}.{\bf v})P_{1} 
\label{Dmunu}
\end{equation}   
and a drift term
\begin{equation}
\eta^{\mu}=-{N\gamma^{2}\over 8} \int d^{2}{\bf r}_{1}
K^{\mu\nu}\delta({\mb\xi}.{\bf v}){\partial P_{1}\over\partial r_{1}^{\nu}},  
\label{etamu}
\end{equation}   
equation (\ref{sun1}) can be rewritten in the more illuminating form:
\begin{equation}
{\partial P\over\partial t}+\langle {\bf V}\rangle \nabla
P={\partial\over\partial r^{\mu}}\biggl \lbrack D^{\mu\nu}{\partial
P\over\partial r^{\nu}}+P\eta^{\mu}\biggr\rbrack
\label{FPgenee}
\end{equation}
similar to the general Fokker-Planck equation (\ref{FP1}). Note, however, that
equation (\ref{FPgenee}) is an integrodifferential equation since the density
probability $P({\bf r},t)$ in ${\bf r}$ at time $t$ depends on the value of the
whole distribution of probability $P({\bf r}_{1},t)$ at the same time by an
integration over ${\bf r}_{1}$. By contrast, the Fokker-Planck equation
(\ref{FP3}) is a differential equation. The usual way to transform an
integrodifferential equation into a differential equation is to make a guess for
the function $P({\bf r}_{1})$ appearing under the integral sign and refine the
guess by successive iterations. In practice we simply make one sensible guess.
Therefore, if we are close to equilibrium, it seems natural to replace the
function $P_{1}$ appearing in the integrals by the Boltzmann distribution 
\begin{equation}
P({\bf r}_{1})=Ae^{-\beta\gamma\psi' ({\bf r}_{1})}.
\label{eq}
\end{equation} 
This corresponds to the ``thermal bath approximation'' of section
\ref{sec_bath}: the vortices have not yet relaxed completely, but when we focus
on the relaxation of a given point vortex (described by $P$) we can consider, in
a first approximation, that the rest of the system (described by $P_{1}$) is at
equilibrium. Within this approximation, the diffusion coefficient and the drift
simplify in:
\begin{equation}
\eta^{\mu}=\beta\gamma D^{\mu\nu}{\partial\psi'\over\partial r^{\nu}}
\label{driftbis}
\end{equation} 
\begin{equation}
D^{\mu\nu}={N\gamma^{2}\over 8}P({\bf r},t)\int K^{\mu\nu}\delta({\mb\xi}.{\bf
v}) d^{2}{\mb\xi},
\label{diffusion}
\end{equation}
where we have made the local approximation. If we assume that the correlation time is short, i.e.
if we replace $\xi^{2}\delta({\mb\xi}.{\bf v})$ by $\tau/\pi^{2}$ (compare
equations (\ref{sun1}) and (\ref{genK4})), we obtain 
\begin{equation}
{\mb\eta}=\beta\gamma D\nabla\psi',
\label{tu1}
\end{equation} 
\begin{equation}
D={\gamma\tau\over 16\pi}\ln N \langle\omega\rangle.
\label{tu2}
\end{equation}
In that case, equation (\ref{FPgenee}) reduces to the Fokker-Planck equation
found in section \ref{sec_relaxation}:
\begin{equation}
{\partial P\over \partial t}+\langle {\bf V}\rangle \nabla P=\nabla(D(\nabla
P+\beta\gamma P\nabla\psi')).
\label{rs}
\end{equation} 
If instead of the Boltzmann distribution (\ref{eq}) we use the Gaussian
distribution (\ref{Gauss}), we get
\begin{equation}
{\partial P\over\partial t}+\langle {\bf V}\rangle\nabla P=\nabla(D(\nabla 
P+\alpha\gamma  P{\bf r})).
\label{FP2einfin}
\end{equation} 
This equation is closely related to the Kramers-Chandrasekhar equation
(\ref{KC}) since the drift and the friction are linear in ${\bf r}$ and ${\bf
v}$ respectively.

The diffusion coefficient (\ref{tu2}) was previously obtained by \cite{csr,rr,chavquasi} using phenomenological
arguments. However, in these studies the correlation time $\tau$ was not
specified. A first determination of $\tau$ was obtained in Ref. \cite{chav98}
but it was  restricted to axisymmetrical or unidirectional flows.  Using
equation (\ref{diffusion}), we can determine the expression of the diffusion
coefficient and the correlation time in the general case. Expanding the velocity
difference ${\bf v}=\langle {\bf V}_{1}\rangle -\langle {\bf V}\rangle$ in a Taylor series in ${\mb\xi}={\bf
r}_{1}-{\bf r}$, we obtain to first order in the expansion   
\begin{equation}
{\mb\xi}.{\bf v}=\Sigma^{\mu\nu}\xi^{\mu}\xi^{\nu},
\label{sat1}
\end{equation}
where 
\begin{equation}
\Sigma^{\mu\nu}={1\over 2}\biggl ({\partial\langle V\rangle^{\mu}\over\partial
r^{\nu}}+{\partial\langle V\rangle^{\nu}\over\partial r^{\mu}}\biggr )
\label{sat2}
\end{equation}
is the stress tensor. It satisfies the property of symmetry
$\Sigma^{\mu\nu}=\Sigma^{\nu\mu}$. Since the flow is divergenceless, we also
have $\Sigma^{xx}+\Sigma^{yy}=0$. This suggest to introduce the notations
$a=\Sigma^{xx}=-\Sigma^{yy}$ and $b=\Sigma^{xy}=\Sigma^{yx}$. In terms of the
stress tensor (\ref{sat2}), the diffusion tensor (\ref{diffusion}) can be
rewritten
\begin{equation}
D^{\mu\nu}={N\gamma^{2}\over 8}P\int
{\xi^{2}\delta^{\mu\nu}-\xi^{\mu}\xi^{\nu}\over
\xi^{2}}\delta(\Sigma^{\mu\nu}\xi^{\mu}\xi^{\nu})  d^{2}{\mb\xi}
\label{sat3}
\end{equation}
This integral can be performed easily by working in a basis where the tensor
$\Sigma^{\mu\nu}$ is anti-diagonal. To that purpose, we seek a tensor $T$ such
that $\Sigma= \tilde T\Sigma' T$ where $\Sigma'$ is anti-diagonal and, by
definition,  $\tilde T^{\mu\nu}=T^{\nu\mu}$. We also impose that $T$ is unitary
so that $\tilde T=T^{-1}$. Then, if we denote by $(M^{11},M^{12},M^{21},M^{22})$
the components of a $2\times 2$ matrix, we find $\Sigma'=(0,b',b',0)$ and
$T=(\alpha,\beta,-\beta,\alpha)$ with $\alpha^{2}+\beta^{2}=1$ and
$b'=b/(\alpha^{2}-\beta^{2})=-a/2\alpha\beta$. From the above results, it is
also clear that $b'^{2}=-Det(\Sigma)=(a^{2}+b^{2})$, where $Det(\Sigma)$ stands
for the determinant of the matrix $\Sigma$. Now, introducing a new system of
coordinates such that $\xi^{'\mu}=T^{\mu\nu}\xi^{\nu}$, or alternatively
$\xi^{\mu}=T^{\nu\mu}\xi^{'\nu}$,  we easily check that the Jacobian of the
transformation ${\mb\xi}\rightarrow {\mb\xi'}$, i.e. the determinant of $T$, is
equal to one. Under these circumstances, the diffusion tensor (\ref{sat3}) can
be written $D=\tilde T D' T$ with 
\begin{equation}
D'^{\mu\nu}={N\gamma^{2}\over 8}P\int
{\xi'^{2}\delta^{\mu\nu}-\xi'^{\mu}\xi'^{\nu}\over \xi'^{2}}\delta
(|\Sigma({\bf r})|\xi'_{1}\xi'_{2})\  d\xi'_{1}d\xi'_{2},
\label{sat4}
\end{equation}
where we have set $|\Sigma({\bf r})|=2\sqrt{-Det(\Sigma)}$. Physically, this
quantity represents the local shear of the flow. The components of the tensor
(\ref{sat4}) can now be determined easily. First of all,
\begin{equation}
D'^{11}={N\gamma^{2}\over 8}{1\over |\Sigma({\bf r})| } P\int {\xi_{2}^{'2}\over
\xi_{1}^{'2}+\xi_{2}^{'2}}\delta (\xi'_{1}\xi'_{2})\ d\xi'_{1}d\xi'_{2}
\label{sat5}
\end{equation}
Setting $\xi_{1}'=\xi\cos\theta$ and $\xi'_{2}=\xi\sin\theta$ where
$\xi=\xi'=|{\bf r}_{1}-{\bf r}|$, we get
\begin{equation}
D'^{11}={N\gamma^{2}\over 8}{1\over |\Sigma({\bf r})| }
P\int_{0}^{+\infty}\xi d\xi\int_{0}^{2\pi}d\theta \sin^{2}\theta\delta
(\xi^{2}\cos\theta\sin\theta),
\label{sat6}
\end{equation}
or, equivalently,
\begin{equation}
D'^{11}={N\gamma^{2}\over 4}{1\over |\Sigma({\bf r})|   }
P\int_{0}^{+\infty}{d\xi\over\xi} \int_{0}^{\pi}d\theta \sin\theta\delta
(\cos\theta).
\label{sat7}
\end{equation}
As explained in Appendix \ref{sec_statistics}, we regularize the logarithmic
divergence by introducing appropriate cut-offs at small and large scales. With
the change of variables $t=\cos\theta$, we obtain 
\begin{equation}
D'^{11}={N\gamma^{2}\over 8}{1\over |\Sigma({\bf r})|   } P\ln N
\int_{-1}^{+1}dt \delta (t)={N\gamma^{2}\over 8}{1\over |\Sigma({\bf r})|  }
P\ln N.
\label{sat8}
\end{equation}
By the same arguments, we find that $D'^{22}=D'^{11}=D$. On the other hands, it follows for reasons of antisymmetry by the transformation
$\xi'_{1}\rightarrow -\xi'_{1}$ that $D'^{12}=D'^{21}=0$. Therefore
$D'^{\mu\nu}=D\delta^{\mu\nu}$ is diagonal in the basis where $\Sigma$ is
anti-diagonal. This remains true in any basis since
$D^{\mu\nu}=T^{\lambda\mu}D'^{\lambda\sigma}T^{\sigma\nu}=D(\tilde T
T)^{\mu\nu}=D\delta^{\mu\nu}$. Therefore, close to equilibrium, the diffusion is isotropic and the
general expression of the diffusion coefficient is
\begin{equation}
D={\gamma\over 8}{1\over |\Sigma({\bf r})|} \ln N \langle\omega\rangle.
\label{sat10}
\end{equation}
Comparing with equation  (\ref{tu2}) we find that the correlation time is given
by
\begin{equation}
\tau={2\pi\over  |\Sigma({\bf r})|   }
\label{sat11}
\end{equation}
The quantity $|\Sigma({\bf r})|=2\sqrt{-Det(\Sigma)}$ plays a fundamental role
in the theory. Clearly, this expression is invariant by a change of referential. For a
unidirectional flow or an axisymmetrical flow, we recover the results of section
\ref{sec_FPeq} and of Ref. \cite{chav98}.

\section{Conclusion}
\label{sec_conclusion}

In this paper, we have provided a systematic derivation of the kinetic equations
of point vortices, applying for the first time the powerful projection operator
technics to this problem. We have described how a cloud of point vortices
relaxes toward the mean field statistical equilibrium, leading to a clustering
into large coherent vortices. In the first part of the paper, we have focused on
the relaxation of a ``test'' vortex in a cloud of background vortices at
statistical equilibrium. The cloud of ``field'' vortices plays the role of a
thermal bath like in other problems of statistical physics. We have shown that
the test vortex undergoes a usual diffusion effect due to random fluctuations
and that it also experiences a {\it systematic drift}. This drift, due to a
polarization of the background vortices by the test vortex, balances the effect
of diffusion at equilibrium, providing a dynamical explanation for the
persistence of clustering. The drift was previously derived with a linear
response theory (Chavanis, 1998), but the diffusion was heuristically introduced
by adding a white noise effect. The present derivation systematically derives
the two effects, diffusion and drift, from the same formalism, and is therefore
more satisfactory. The diffusion derived here turns out to be influenced by long
time correlations, so it is more complex than the usual white noise effect. When
memory effects are ignored, we obtain a Fokker-Planck equation for the evolution
of the one-particle distribution function. This Fokker-Planck equation can also
be derived from a  phenomenological Maximum Entropy Production Principle
\cite{rs}. This shows that the structure of this equation is influenced
more by thermodynamics (the first and second principles) than by the precise
microscopic model. However, our  systematic procedure starting directly from the
Liouville equation  provides  a {\it justification} for this thermodynamical
approach and specifies its range a validity. It also allows to determine
explicitly the value of the diffusion coefficient which was left unspecified by
the Maximum Entropy Production Principle. All these results could be tested
numerically by introducing a test vortex in ``sea'' of vortices at statistical
equilibrium and by solving the Kirchhoff-Hamilton equations of motion.   

In the second part of the paper, we have attempted to describe the evolution of
the whole system of vortices far from equilibrium. We have obtained a new
kinetic equation (\ref{new1}) which incorporates a delocalization in space and
time. This is therefore a non Markovian integrodifferential equation. A similar
equation also occurs in the quasilinear theory of the 2D Euler equation
(Chavanis, 2000). Within some approximation, it is possible to carry out the
time integration explicitly and this yields a simpler equation (\ref{sun1})
which only conserves a delocalization in space. This equation respects all the
conservation laws of the point vortex system and satisfies a H-theorem. The
relaxation is due to a condition of resonance between distant vortices.  If the
system is sufficiently ``resonant'', it will reach a maximum entropy state
described by the Boltzmann distribution. However, if there are not enough
resonances the evolution may stop on metastable state. Only non trivial
correlations between vortices (for example three body collisions) can  unfreeze
the system and induce further evolution. These correlations are not taken into
account in the present analysis although the projection operator formalism might 
still apply. It would be necessary to modify the factorization hypothesis 
(\ref{decorrsys}) so as to account for two-body or three-body correlation 
functions. It is possible, however, that these correlations develop on a much 
longer time scale so it is not yet clear whether they are physically relevant.
In any case, the approximations made in the present paper are a first step 
towards a rational kinetic theory of point vortices.

It is also possible that a system of point vortices undergoes a form of
``violent relaxation'' in its early stage. For short time scales, the 
correlations between point
vortices have not yet developed and the average vorticity is solution of the 2D
Euler equation. When the initial condition is far from equilibrium, it is
well-known that the 2D Euler equation develops a complicated mixing process
leading to the formation of an organized state (on a coarse-grained scale). 
This relaxation is quite rapid,
of the order of the dynamical time $t_{D}$, and the resulting equilibrium state
is predicted to be a complicated superposition of Fermi-Dirac distributions
respecting all the constraints of the Euler equation \cite{miller,rseq}. On
longer time scales, of the order of ${N\over \ln N}t_{D}$, the correlations
between point vortices develop and the system undergoes another form of
relaxation, much slower. This relaxation is towards the Boltzmann distribution
derived by \cite{jm} Joyce \& Montgomery and \cite{lp} Lundgren \& Pointin which
is the true equilibrium state for a system of point vortices. The first type of
relaxation, by vorticity mixing,  was described in Ref. \cite{chavquasi} using a
quasilinear theory of the 2D Euler equation. The exclusion principle leading to
the Fermi-Dirac statistics was explicitly shown as well as a $H$ theorem. The second type of relaxation,
due to discrete interactions between point vortices, was the object of the
present paper. 

It is noteworthy that a similar distinction occurs in the context of stellar
systems \cite{csr}. Indeed, the relaxation of stars is a two stage process. For
short time scales $\sim t_{D}$, the encounters between stars can be neglected
and the distribution function is solution of the Vlasov equation (analogous to
the 2D Euler equation). If the system is initially far from mechanical
equilibrium, it will experience a ``violent relaxation'' towards a virialized
state. This equilibrium is predicted to be a superposition of Fermi-Dirac
statistics \cite{lb,mnrasCS}, like for the 2D Euler equation. Then, on a longer time
scale, of the order of ${N\over \ln N}t_{D}$, the encounters between stars
cannot be ignored anymore and will deviate the stars from their unperturbed
trajectories. This collisional relaxation is usually described by a Landau or
Fokker-Planck equation that converges towards the Maxwell-Boltzmann distribution
at equilibrium. On even longer time scales, three-body encounters leading to the
formation of binaries induce a slow evolution of the system \cite{bt}. Three
body encounters (involving vortices of different sign) are also relevant in two
dimensional turbulence and lead to the formation of vortex ``pairs'' \cite{sc}.
The analogy with ``binary stars'' is interesting to note. It would be important
to test these ideas with numerical simulations of point vortices or stars. The
situation is difficult in the stellar context  because a maximum entropy state
does not always exist. Indeed, the system can collapse and overheat: this is the
so-called ``gravothermal catastrophe'' \cite{lbw}. This problem does not occur
for point vortices and it should be possible to evidence the two (or more)
successive equilibria more properly. An advantage of point vortices with respect
to stars is the lower dimensionality of space ($D=2$ instead of $D=3$, or $D=6$ in phase space) that
should make numerical simulations easier.

\section{Acknowledgments}
\label{sec_ack}

I would like to thank J. Sommeria, P. Newton, I. Mezic, A. Provenzale
and F. Bouchet 
for their interest in this study.
This research was supported in part by the National Science Foundation under
Grant No. PHY94-07194.

\newpage
\appendix

\section{The statistics of velocity fluctuations in an inhomogeneous medium  }
\label{sec_statistics}

In this Appendix, we study the statistics  of velocity fluctuations produced by
an inhomogeneous distribution of point vortices. This study extends the
calculations of \cite{jimenez,min,cs} for a uniform medium and provides a simple
framework to understand the logarithmic divergence of the diffusion coefficient.
Let us consider a collection of $N$ point vortices randomly distributed in a
disk of radius $R$ with an average density $n({\bf r})$. The velocity ${\bf V}$
occuring at a given location ${\bf r}$ of the flow is the sum of the velocities
${\mb\Phi}_{i}$ ($i=1,...,N$) produced by the N vortices:
\begin{equation}
{\bf V}=\sum_{i=1}^{N} {\mb\Phi}_{i},
\label{S1}
\end{equation} 
\begin{equation}
{\bf \Phi}_{i}=-{\gamma\over 2\pi}{({\bf r}_{i}-{\bf r})_{\perp}\over |{\bf
r}_{i}-{\bf r}|^{2}}.
\label{S2}
\end{equation} 
Following a procedure similar to that adopted in Ref. \cite{cs}, the velocity
distribution can be expressed as
\begin{equation}
W_{N}({\bf V})={1\over 4\pi^{2}}\int A_{N}({\mb \rho})e^{-i{\mb\rho}{\bf
V}}d^{2}{\mb\rho}
\label{S3}
\end{equation} 
with
\begin{equation}
A_{N}({\mb\rho})=\biggl (\int e^{i{\mb\rho}{\mb \Phi}}P({\bf r}_{1}) d^{2}{\bf
r}_{1}\biggr )^{N}.
\label{S4}
\end{equation}
Here,  $P({\bf r}_{1})$ denotes the probability of occurence of a point vortex
in ${\bf r}_{1}$ and by definition
\begin{equation}
{\mb \Phi}=-{\gamma\over 2\pi}{({\bf r}_{1}-{\bf r})_{\perp}\over |{\bf
r}_{1}-{\bf r}|^{2}}.
\label{S5}
\end{equation} 
Introducing explicitly the vortex density $n({\bf r}_{1})=NP({\bf r}_{1})$, we
have
\begin{equation}
A_{N}({\mb\rho})=\biggl ({1\over N}\int n({\bf r}_{1}) e^{i{\mb\rho}{\mb \Phi}}
d^{2}{\bf r}_{1}\biggr )^{N}.
\label{S6}
\end{equation} 
Since 
\begin{equation}
\int n({\bf r}_{1})d^{2}{\bf r}_{1}=N,
\label{S7}
\end{equation} 
we can rewrite our expression for $A_{N}({\mb\rho})$ in the form
\begin{equation}
A_{N}({\mb\rho})=\biggl (1-{1\over N}\int n({\bf r}_{1}) (1-e^{i{\mb\rho}{\mb
\Phi}}) d^{2}{\bf r}_{1}\biggr )^{N}.
\label{S8}
\end{equation} 
In the limit of large $N$, $R$ with fixed  $n({\bf r})$, we can approximate the
foregoing expression by
\begin{equation}
A({\mb\rho})=e^{-C({\mb\rho})}
\label{S9}
\end{equation} 
with
\begin{equation}
C({\mb\rho})=\int n({\bf r}_{1}) (1-e^{i{\mb\rho}{\mb \Phi}})d^{2}{\bf r}_{1}.
\label{S10}
\end{equation} 
Separating  the real and imaginary parts of $C({\mb\rho})$, we obtain
\begin{eqnarray}
C({\mb\rho})=C_{1}({\mb\rho})-i C_{2}({\mb\rho})\nonumber\\ 
=\int n({\bf r}_{1})
(1-\cos({\mb\rho}{\mb \Phi}))d^{2}{\bf r}_{1}-i\int n({\bf r}_{1}) 
\sin({\mb\rho}{\mb \Phi}) d^{2}{\bf r}_{1}.
\label{S11}
\end{eqnarray} 
In the first integral, we find it convenient to introduce the relative
separation ${\mb\xi}={\bf r}_{1}-{\bf r}$ in terms of which
\begin{equation}
C_{1}({\mb\rho})=\int n({\bf r}+{\mb\xi}) (1-\cos({\mb\rho}{\mb
\Phi}))d^{2}{\mb\xi}
\label{S12}
\end{equation} 
with
\begin{equation}
{\bf \Phi}=-{\gamma\over 2\pi}{{\mb\xi}_{\perp}\over {\xi}^{2}}.
\label{S13}
\end{equation} 
In Ref. \cite{cs}, it was found that an important contribution to the velocity
fluctuations comes from the nearest neigbour. This justifies to make the local
approximation $n({\bf r}+{\mb\xi})\simeq n({\bf r})$ in equation (\ref{S12}). In
this approximation
\begin{equation}
C_{1}({\mb\rho})=n({\bf r})\int  (1-\cos({\mb\rho}{\mb \Phi}))d^{2}{\mb\xi}.
\label{S14}
\end{equation} 
This quantity is closely related to the function $C({\mb\rho})$ evaluated in
Ref. \cite{cs} for a distribution of vortices with uniform density $n$. The only
difference is the presence of the local density  $n({\bf r})$  in place of $n$.
Therefore, we can infer directly that
\begin{equation}
C_{1}({\mb\rho})={n({\bf r})\gamma^{2}\over 8\pi}\ln\biggl ({2\pi R\over
\gamma\rho}\biggr )\rho^{2}.
\label{S15}
\end{equation} 
In fact, the local approximation is only marginally valid because, as discussed
in Ref. \cite{cs}, the contribution of the nearest neighbor is precisely of the
same order of magnitude as the contribution of the  rest of the system. This
results in a logarithmic divergence in (\ref{S15}) due to the weak collective
behaviour of the system.

In the second integral appearing in (\ref{S11}), the contribution from proximate
vortices  vanishes by symmetry. As a result, the integral is dominated by large
values of $|{\bf r}_{1}|$ or, equivalently, by small values of $|{\mb\Phi}|$. We
can therefore make the approximation $\sin( {\mb\rho}{\mb\Phi})\simeq
{\mb\rho}{\mb\Phi}$ and write 
\begin{equation}
C_{2}({\mb\rho})={\mb\rho}\int n({\bf r}_{1}){\mb\Phi} d^{2}{\bf r}_{1}. 
\label{S16}
\end{equation}
In the integral, we recognize the mean-field velocity created in ${\bf r}$ by
the average distribution  of vortices:
\begin{equation}
\langle {\bf V}\rangle({\bf r}) =\int n({\bf r}_{1}){\mb\Phi}d^{2}{\bf r}_{1}.
\label{S17}
\end{equation}
Hence,
\begin{equation}
C_{2}({\mb\rho})={\mb\rho}\langle {\bf V}\rangle({\bf r}).
\label{S18}
\end{equation}

Substituting the explicit expression for $C({\mb\rho})$ in equation (\ref{S9}),
we obtain
\begin{equation}
A({\mb\rho})=e^{-{n({\bf r})\gamma^{2}\over 8\pi}\ln ({2\pi R\over
\gamma\rho})\rho^{2}+i{\mb\rho}\langle {\bf V}\rangle({\bf r})}.
\label{S19}
\end{equation}
Therefore, the velocity distribution in ${\bf r}$ can be written quite generally
\begin{equation}
W({\bf V})={1\over 4\pi^{2}}\int e^{-{n({\bf r})\gamma^{2}\over 8\pi}\ln ({2\pi
R\over \gamma\rho})\rho^{2}}e^{-i{\mb\rho}({\bf V}-\langle {\bf V}\rangle ({\bf
r}))}d^{2}{\mb\rho}.
\label{S20}
\end{equation}
This is the same distribution as for a uniform distribution of vortices except
that the constant density $n$ has been replaced by the local density  $n({\bf
r})$ and that the distribution is for the fluctuating velocity ${\mb{\cal
V}}={\bf V}-\langle {\bf V}\rangle({\bf r})$. Repeating the calculations of Ref.
\cite{cs}, the velocity p.d.f is explicitly given by:
\begin{equation}
W({\mb{\cal V}})= {4\over  {n({\bf r})\gamma^{2}} \ln N} e^{-{4\pi\over n({\bf
r}) {\gamma^{2}}\ln N} {\cal V}^{2}}\quad ({\cal V}\lesssim {\cal V}_{crit}(N))
\label{S21}
\end{equation} 
\begin{equation}
W({\mb{\cal V}})\sim {n({\bf r})\gamma^{2}\over 4 \pi^{2}{\cal V}^{4}}\quad
({\cal V}\gtrsim {\cal V}_{crit}(N))
\label{S22}
\end{equation}
where 
\begin{equation}
{\cal V}_{crit}(N)\sim \biggl ({n({\bf r})\gamma^{2}\over 4\pi}\ln N\biggr
)^{1/2}\ln^{1/2}(\ln N).
\label{S23}
\end{equation}
This distribution lies at the frontier between Gaussian and L\'evy laws: the
core of the distribution is Gaussian while the tail decays algebraically like
for L\'evy laws. This is because the variance of the individual velocities
(\ref{S2}) diverges logarithmically so the central limit theorem is only
marginally applicable. For that reason we have proposed to call this
distribution the ``marginal Gaussian distribution'' \cite{cs2}. In the strict
mathematical limit $N\rightarrow +\infty$, the transition between the two
regimes is rejected to infinity and  the velocity p.d.f. is purely Gaussian
\cite{min}. However,  the convergence towards this Gaussian distribution is so
slow that in practical applications it is never reached: the algebraic tail
always remain \cite{weiss}.

According to equations (\ref{S1}) and  (\ref{S2}), the variance of the velocity
can be written
\begin{equation}
\langle {\cal V}^{2}\rangle = \int_{|{\mb \xi}|=0}^{R} n({\bf
r}+{\mb\xi}){\gamma^{2}\over 4\pi^{2}\xi^{2}}d^{2}{\mb \xi}=n({\bf r})
\int_{0}^{R}{\gamma^{2}\over 4\pi^{2}\xi^{2}}2\pi\xi d\xi,
\label{S25}
\end{equation}  
where we have made the local approximation in the second equality. This quantity
diverges logarithmically at both small and large vortex separations. The
divergence at small separations is a failure of our model which ignores
correlations between vortices. In fact, when two vortices approach each other
they can form a pair, as discussed in Ref. \cite{weiss}, 
and our mean field theory
clearly breaks down. We shall account heuristically for this failure by
introducing a cut-off at some minimum distance $d_{pair}\sim (\pi n\ln
N)^{-1/2}$ \cite{cs}. The divergence at large separations is due to the
unshielded nature of the interaction potential. It is therefore natural to cut
the integral at $R$, the typical size of the system. With this regularization,
we obtain 
\begin{equation}
\langle {\cal V}^{2}\rangle = {n({\bf r})\gamma^{2}\over 2\pi}\ln \biggl
({R\over d}\biggr )= {n({\bf r})\gamma^{2}\over 4\pi}\ln N. 
\label{S27}
\end{equation} 
Since the divergence is weak (logarithmic), the result does not depend crucially
on the precise value of the cutoffs. The same expression for the variance can
also be obtained from the formula 
\begin{equation}
\langle {\cal V}^{2}\rangle=\int W({\cal V}){\cal V}^{2}2\pi {\cal V} 
d{\cal V}
\label{S24}
\end{equation}
if we introduce a cut-off at large velocities and use equations  (\ref{S21})
(\ref{S22}) (\ref{S23}) [to sufficient accuracy, we just need considering
equation (\ref{S21})].

The diffusion coefficient can be expressed in terms of the variance (\ref{S27})
and the typical correlation time $\tau$ by \cite{cs}:
\begin{equation}
D={1\over 4} \tau \langle {\cal V}^{2}\rangle.
\label{S28}
\end{equation}
Using equation (\ref{S27}) and the relation $\langle\omega\rangle=n\gamma$, we
obtain
\begin{equation}
D={\gamma\tau\over 16\pi}\ln N\langle\omega\rangle
\label{S28new}
\end{equation}
in agreement with our result (\ref{tu2}). The correlation time is more difficult
to evaluate. If we ignore the differential rotation of the vortices, the
calculations of Ref. \cite{cs} are directly applicable and lead to the
expression
\begin{equation}
\tau\sim{1\over \langle \omega\rangle ({\bf r})\sqrt{\ln N}}.
\label{S29}
\end{equation}
This is the typical time needed by a vortex to cross the interparticle distance
$d\sim 1/\sqrt{n({\bf r})}$ with the velocity $\sqrt{ \langle {\cal
V}^{2}\rangle}$. With this approximation, the diffusion coefficient is given by
equation (\ref{conc5}). On the other hand, if we consider that the vortices are
transported by the equilibrium flow and evaluate the diffusion coefficient with
the Kubo formula (see Appendix \ref{sec_diffcoeff}), we find that the
correlation time is related to the local shear by equation (\ref{sat11}), i.e.:
\begin{equation}
\tau={2\pi\over |\Sigma({\bf r})|}.
\label{S30}
\end{equation}  
Physically, it corresponds to the time needed by two vortices with relative
velocity $\Sigma d$ to be stretched by the shear on a distance $\sim d$. This
approximation breaks up, however, when the shear is weak. In that case $\tau$ is
given by equation (\ref{S29}) obtained when only the dispersion of the vortices
is considered. Clearly, a general formula should take into account
simultaneously the effect of the shear and the dispersion of the particles.

\section{ The Kubo formula}
\label{sec_diffcoeff}

Let us consider the diffusion of a test vortex in a ``sea" of field vortices
described by the equilibrium distribution 
\begin{equation}
\mu_{eq}({\bf r}_{1},...,{\bf r}_{N})=\prod_{i=1}^{N}P_{eq}({\bf r}_{i}).
\label{Amueq}
\end{equation}
The general form of the diffusion coefficient, as defined by equation
(\ref{FP1}), writes
\begin{equation}
D^{\mu\nu}={\langle \Delta r^{\mu}\Delta r^{\nu}\rangle\over 2\Delta t}.
\label{AD}
\end{equation}
Now, the net displacement of the test vortex produced by the fluctuations of the
velocity between $t$ and $t+\Delta t$ is given by
\begin{equation}
\Delta {\bf r}=\int_{t}^{t+\Delta t}{\mb {\cal V}}(t')dt'.
\label{ADeltar}
\end{equation}
Substituting explicitly for $\Delta {\bf r}$ from equation (\ref{ADeltar}) in
equation (\ref{AD}), we have 
\begin{equation}
D^{\mu\nu}={1\over 2\Delta t}\int_{0}^{\Delta t}dt'\int_{0}^{\Delta
t}dt''\langle {\cal V}^{\mu}(t+t'){\cal V}^{\nu}(t+t'')\rangle_{eq},
\label{AD1}
\end{equation}  
where $\langle . \rangle_{eq}$ denotes the average with respect to the
equilibrium distribution (\ref{Amueq}). Since the correlation function appearing in the integral only depends on the
time difference $|t''-t'|$, we also have 
\begin{equation}
D^{\mu\nu}={1\over \Delta t}\int_{0}^{\Delta t}dt'\int_{0}^{t'}dt''\langle {\cal
V}^{\mu}(t+t'){\cal V}^{\nu}(t+t'')\rangle_{eq}
\label{AD2}
\end{equation}
or, alternatively,
\begin{equation}
D^{\mu\nu}={1\over \Delta t}\int_{0}^{\Delta t}dt'\int_{0}^{t'}dt''\langle {\cal
V}^{\mu}(t){\cal V}^{\nu}(t+t''-t')\rangle_{eq}. 
\label{AD3}
\end{equation} 
With the change of variables $\tau=t'-t''$, we obtain successively
\begin{eqnarray}
D^{\mu\nu}={1\over \Delta t}\int_{0}^{\Delta t}dt'\int_{0}^{t'}d\tau\langle
{\cal V}^{\mu}(t){\cal V}^{\nu}(t-\tau)\rangle_{eq}\nonumber\\
 ={1\over \Delta
t}\int_{0}^{\Delta t}d\tau\int_{\tau}^{\Delta t}dt'\langle {\cal
V}^{\mu}(t){\cal V}^{\nu}(t-\tau)\rangle_{eq}\nonumber\\
 ={1\over \Delta
t}\int_{0}^{\Delta t}d\tau\langle {\cal V}^{\mu}(t){\cal
V}^{\nu}(t-\tau)\rangle_{eq}   (\Delta t-\tau).
\label{AD4}
\end{eqnarray} 
If the correlation function $\langle {\cal V}^{\mu}(t){\cal
V}^{\nu}(t-\tau)\rangle_{eq}$    decays more rapidly than $\tau^{-1}$, we can
take the limit $\Delta t\rightarrow +\infty$ to finally obtain
\begin{equation}
D^{\mu\nu}=\int_{0}^{+\infty}\langle {\cal V}^{\mu}(t){\cal
V}^{\nu}(t-\tau)\rangle_{eq} d\tau.
\label{AD5}
\end{equation} 
This is the Kubo formula for our problem. Remembering that ${\mb{\cal V}}$
denotes the fluctuation of the total velocity:
\begin{equation}
 {\mb{\cal V}}(t)={\bf V}(t)-\langle {\bf V}(t)\rangle_{eq},
\label{Afluct}
\end{equation} 
we find that
\begin{equation}
\langle {\cal V}^{\mu}(t){\cal V}^{\nu}(t-\tau)\rangle_{eq}=\langle
{V}^{\mu}(t){V}^{\nu}(t-\tau)\rangle_{eq}-\langle
{V}^{\mu}(t)\rangle_{eq}\langle {V}^{\nu}(t-\tau)\rangle_{eq}.
\label{Acorr1}
\end{equation} 
Now, the first quantity in bracket can be written explicitly
\begin{eqnarray}
\langle
{V}^{\mu}(t){V}^{\nu}(t-\tau)\rangle_{eq}=\sum_{i=1}^{N}\sum_{j=1}^{N}\int
V^{\mu}(i\rightarrow 0,t)V^{\nu}(j\rightarrow 0,t-\tau)\mu_{eq}(\lbrace{\bf
r}_{k}\rbrace)\prod_{k=1}^{N} d^{2}{\bf r}_{k}\nonumber\\
=\sum_{i=1}^{N}\sum_{j\neq i}\int V^{\mu}(i\rightarrow 0,t)V^{\nu}(j\rightarrow
0,t-\tau)\mu_{eq}(\lbrace{\bf r}_{k}\rbrace)\prod_{k=1}^{N}  d^{2}{\bf
r}_{k}\nonumber\\ 
+\sum_{i=1}^{N}\int V^{\mu}(i\rightarrow
0,t)V^{\nu}(i\rightarrow 0,t-\tau)\mu_{eq}(\lbrace{\bf
r}_{k}\rbrace)\prod_{k=1}^{N} d^{2}{\bf r}_{k}\nonumber\\
 =N(N-1)\langle
V^{\mu}(1\rightarrow 0,t)\rangle_{eq}\langle V^{\nu}(1\rightarrow
0,t-\tau)\rangle_{eq}\nonumber\\
 +N\int  V^{\mu}(1\rightarrow 0,t)
V^{\nu}(1\rightarrow 0,t-\tau)P_{eq}({\bf r}_{1})d^{2}{\bf r}_{1}. 
\label{Acorr2}
\end{eqnarray} 
For large $N$, we can make the approximation $N(N-1)\simeq N^{2}$. Using
(\ref{Acorr1}) and (\ref{Acorr2}) the correlation function can be put in the
form
\begin{equation}
\langle {\cal V}^{\mu}(t){\cal V}^{\nu}(t-\tau)\rangle_{eq}=N\int
V^{\mu}(1\rightarrow 0,t) V^{\nu}(1\rightarrow 0,t-\tau)P_{eq}({\bf
r}_{1})d^{2}{\bf r}_{1}\label{Acorr3}.
\end{equation} 
Since the integral is dominated by interactions involving relatively close
vortices, we can make the local approximation:
\begin{equation}
\langle {\cal V}^{\mu}(t){\cal V}^{\nu}(t-\tau)\rangle_{eq}=N\int
V^{\mu}(1\rightarrow 0,t) V^{\nu}(1\rightarrow 0,t-\tau)P_{eq}({\bf r})d^{2}{\bf
r}_{1}.
\label{Acorr4}
\end{equation} 
The expression for the diffusion coefficient then becomes
\begin{equation}
D^{\mu\nu}=N\int_{0}^{+\infty}d\tau \int  V^{\mu}(1\rightarrow 0,t)
V^{\nu}(1\rightarrow 0,t-\tau)P_{eq}({\bf r})d^{2}{\bf r}_{1}. 
\label{AD6}
\end{equation} 
For sake of brevity, we shall denote the velocity correlation function by
\begin{equation}
C^{\mu\nu}(\tau)\equiv \langle {\cal V}^{\mu}(t){\cal
V}^{\nu}(t-\tau)\rangle_{eq}=N \int  V^{\mu}(1\rightarrow 0,t)
V^{\nu}(1\rightarrow 0,t-\tau)P_{eq}({\bf r})d^{2}{\bf r}_{1}. 
\label{Acorr5}
\end{equation} 
Therefore, the Kubo formula takes the form
\begin{equation}
D^{\mu\nu}=\int_{0}^{+\infty}C^{\mu\nu}(\tau)d\tau 
\label{AD7}
\end{equation} 
More generally, we have 
\begin{equation}
\langle\Delta r^{\mu}\Delta r^{\nu}\rangle=2\int_{0}^{\Delta t}
C^{\mu\nu}(\tau)(\Delta t-\tau)d\tau.
\label{AD7new}
\end{equation}
These quantities are calculated explicitly in Appendix \ref{sec_expcal} in the
case of simple flows.

\section{The calculation of the diffusion coefficient}
\label{sec_expcal}

In this Appendix, we calculate the Kubo integral using an approximation in which
the point vortices  follow the streamlines of the equilibrium flow.

\subsection{Unidirectional flow}
\label{sec_uni}

We shall first calculate the velocity correlation function (\ref{Acorr5}) and
the diffusion coefficient (\ref{AD6}) in the case of an unidirectional
equilibrium flow.  The trajectory of a fluid particle advected by this flow is
simply:
\begin{eqnarray}
y(t-\tau)=y(t),
\label{uni1}
\end{eqnarray}  
\begin{eqnarray}
x(t-\tau)=x(t)-\langle V\rangle_{eq}(y)\tau.
\label{uni2}
\end{eqnarray}  
According to section \ref{sec_uniuni}, we are particularly interested by the
$yy$ component (\ref{yto}) of the velocity correlation function. Explicitly, it
has the form:
\begin{eqnarray}
C(\tau)={N\gamma^{2}\over 4\pi^{2}} \int dx_{1}dy_{1}{x_{1}-x\over
(x_{1}-x)^{2}+(y_{1}-y)^{2}}(t){x_{1}-x \over
(x_{1}-x)^{2}+(y_{1}-y)^{2}}(t-\tau)P_{eq}(y),
\label{uni3}
\end{eqnarray}  
where we have used equation (\ref{Vji}). The second term involves the quantity
\begin{eqnarray}
(x_{1}-x)(t-\tau)=x_{1}-x+(\langle V\rangle_{eq}(y_{1})-\langle V\rangle_{eq}
(y))\tau.
\label{uni4}
\end{eqnarray} 
In the local approximation, we can expand the velocity difference in a Taylor
series in $y_{1}-y$. To first order, we have
\begin{eqnarray}
\langle V\rangle_{eq}(y_{1})-\langle V\rangle_{eq}(y)\simeq -\Sigma(y)(y_{1}-y),
\label{uni5}
\end{eqnarray} 
where $\Sigma(y)$ is the local shear of the flow (\ref{uni6}).
Introducing the variables $X\equiv x_{1}-x$ and $Y\equiv y_{1}-y$, we obtain:
\begin{eqnarray}
C(\tau)={N\gamma^{2}\over 4\pi^{2}}P_{eq}(y) \int dXdY {X\over X^{2}+Y^{2}}\
{X+\Sigma(y) Y\tau\over (X+\Sigma(y) Y\tau)^{2}+Y^{2}}.
\label{uni7}
\end{eqnarray}  
The integration over $X$ can be performed easily since the integrand is just a
rational function of polynomials. After straightforward calculations, we find:
\begin{eqnarray}
C(\tau)={N\gamma^{2}\over 4\pi}P_{eq}(y){1\over 1+{1\over
4}\Sigma^{2}(y)\tau^{2}} \int_{0}^{+\infty}  {dY\over Y}. 
\label{uni8}
\end{eqnarray}  
The integral over $Y$ diverges logarithmically for both small and large $Y$. The
reason for this divergence has been given in Ref. \cite{cs} and in Appendix
\ref{sec_statistics}. Introducing two cut-offs at scales $d$ and $R$, and noting
that $\ln (R/d)\sim {1\over 2}\ln N$, we finally obtain equation (\ref{uni9}).
For $\tau\rightarrow +\infty$, the correlation function decreases like
$\tau^{-2}$. This is a slow decay but still the diffusion coefficient
(\ref{AD7}) converges. Using
\begin{eqnarray}
\int_{0}^{t} C(\tau) d\tau={N\gamma^{2}\over 4\pi}{\ln N\over |\Sigma(y)|}
\arctan\biggl ({1\over 2}|\Sigma(y)|t\biggr )P_{eq}(y)
\label{uni10}
\end{eqnarray} 
and taking the limit $t\rightarrow +\infty$, we find equation (\ref{uni11}).
More generally, using (\ref{AD7new}) and (\ref{uni10}), we have
\begin{eqnarray}
\langle (\Delta y)^{2}\rangle={N\gamma^{2}\over 2\pi}{\ln N\over
|\Sigma(y)|}P_{eq}(y)\biggl\lbrack  \arctan\biggl ({1\over 2}|\Sigma(y)|\Delta
t\biggr )\Delta t-{1\over |\Sigma(y)|}\ln\biggl (1+{1\over
4}\Sigma(y)^{2}(\Delta t)^{2}\biggr )\biggr\rbrack.
\label{uni10new1}
\end{eqnarray} 
For $\Delta t\rightarrow 0$ (ballistic motion): 
\begin{eqnarray}
\langle (\Delta y)^{2}\rangle={N\gamma^{2}\over 8\pi}{\ln N}P_{eq}(y)(\Delta
t)^{2}={1\over 2}\langle {\cal V}^{2}\rangle (\Delta t)^{2}
\label{uni10new2}
\end{eqnarray} 
and for $\Delta t\rightarrow +\infty$ (diffusive motion): 
\begin{eqnarray}
\langle (\Delta y)^{2}\rangle={N\gamma^{2}\over 4}{\ln N\over
|\Sigma(y)|}P_{eq}(y)\Delta t.
\label{uni10new3}
\end{eqnarray}

\subsection{Axisymmetrical  flow}
\label{sec_axi}

In an axisymmetrical flow, the trajectory of a fluid particle takes the simple
form:
\begin{eqnarray}
r(t-\tau)=r(t),
\label{axi1}
\end{eqnarray}  
\begin{eqnarray}
\theta(t-\tau)=\theta(t)-{\langle V\rangle_{eq}(r)\over r}\tau.
\label{axi2}
\end{eqnarray}  
As indicated in section \ref{sec_axiaxi}, we are particularly interested by the
$r(t)r(t-\tau)$ component (\ref{fgr}) of the correlation function. Let us
introduce the separation $\delta {\bf r}\equiv {\bf r}_{1}-{\bf r}$ between the
field vortex $1$ and the test vortex. In the local approximation,  $\delta {\bf
r}$ can be considered as a small quantity. Therefore we can write:
\begin{eqnarray}
\delta {\bf r}=r\delta\theta {\bf e}_{\theta}+\delta r{\bf e}_{r}\equiv X{\bf
e}_{\theta}+Y{\bf e}_{r},
\label{axi4}
\end{eqnarray}
\begin{eqnarray}
d^{2}{\bf r}_{1}=d^{2}(\delta {\bf r})=dXdY.
\label{axi5}
\end{eqnarray}
With these notations, the correlation function (\ref{fgr}) can be rewritten:
\begin{eqnarray}
C(\tau)={N\gamma^{2}\over 4\pi^{2}}P_{eq}(r)\int dXdY {X\over
X^{2}+Y^{2}}(t){X\over X^{2}+Y^{2}}({t-\tau}).
\label{axi6}
\end{eqnarray}
Now,
\begin{eqnarray}
Y(t-\tau)=\delta r(t-\tau)=r_{1}(t-\tau)-r(t-\tau)
=r_{1}(t)-r(t)=Y(t)=Y
\label{axi7}
\end{eqnarray}
and
\begin{eqnarray}
X(t-\tau)=r(t-\tau)\delta\theta(t-\tau)=r(t-\tau)\biggl
(\theta_{1}(t-\tau)-\theta (t-\tau)\biggr )\nonumber\\
 = r\biggl
(\theta_{1}(t)-\theta(t)-\biggl ({\langle V\rangle_{eq}(r_{1})\over
r_{1}}-{\langle V\rangle_{eq}(r)\over r}\biggr )\tau\biggr ).
\label{axi77}
\end{eqnarray}
In the local approximation, we can expand the last term in equation
(\ref{axi77}) in a Taylor series in $r_{1}-r$. This yields
\begin{eqnarray}
X(t-\tau)=r (\theta_{1}(t)-\theta (t))-r {d\over dr}\biggl ({\langle
V\rangle_{eq}(r)\over r}\biggr ) (r_{1}-r) \tau =X-\Sigma(r)Y\tau.
\label{axi8}
\end{eqnarray}
where $\Sigma(r)$ is the local shear of the flow (\ref{axi9}). Substituting
equations (\ref{axi7}) and  (\ref{axi8}) in equation (\ref{axi6}), we get
\begin{eqnarray}
C(\tau)={N\gamma^{2}\over 4\pi^{2}}P_{eq}(r)\int dXdY {X\over
X^{2}+Y^{2}}{X-\Sigma(r) Y \tau\over (X-\Sigma(r) Y \tau)^{2}+Y^{2}}.
\label{axi10}
\end{eqnarray}
This integral is similar to (\ref{uni7}), so we directly obtain equations
(\ref{axi11}) and (\ref{axi12}).

\section{Conservation laws satisfied by the generalized kinetic equation}
\label{sec_conservation}

In this section, we prove some general properties satisfied by equation
(\ref{new1}). Note first that it can be written
\begin{eqnarray}
{\partial P\over\partial t}+\langle {\bf V}\rangle\nabla P=-\nabla . {\bf J},
\label{c1}
\end{eqnarray}
where 
\begin{eqnarray}
{\bf J}=-N\int_{0}^{t}d\tau\int d^{2}{\bf r}_{1}{\bf V}(1\rightarrow
0)_{t}\biggl\lbrace {\bf V}(1\rightarrow 0)P_{1}\nabla P+{\bf V}(0\rightarrow
1)P\nabla P_{1}\biggr\rbrace_{t-\tau}
\label{c2}
\end{eqnarray}
is the diffusion current. It is clear at first sight that equation (\ref{c1})
conserves the total circulation $\Gamma=\int\langle\omega\rangle d^{2}{\bf r}$
provided that ${\bf J}.\hat{\bf n}=0$ on the boundary. We now prove the
conservation of other integral constraints depending on the domain shape.

(i) In a circular and in an infinite domain, the angular momentum  
defined by
\begin{eqnarray}
L=\int \langle\omega\rangle r^{2}d^{2}{\bf r}
\label{c3}
\end{eqnarray}
must be conserved. Taking the time derivative of equation
(\ref{c3}), substituting for (\ref{c1}) and remembering that $L$ is conserved by
the advective term, we get
\begin{eqnarray}
\dot L=2N\gamma\int {\bf J}. {\bf r}d^{2}{\bf r}.
\label{c4}
\end{eqnarray}
Substituting explicitly for the diffusion current (\ref{c2}) in equation
(\ref{c4}), we obtain
\begin{eqnarray}
\dot L=-2N^{2}\gamma\int_{0}^{t}d\tau\int d^{2}{\bf r}d^{2}{\bf r}_{1}\lbrack
{\bf r}.{\bf V}(1\rightarrow 0)\rbrack_{t} \biggl\lbrace {\bf V}(1\rightarrow
0)P_{1}\nabla P+{\bf V}(0\rightarrow 1)P\nabla P_{1}\biggr\rbrace_{t-\tau}.
\label{c5}
\end{eqnarray}
Permuting the dummy variables ${\bf r}$ and ${\bf r}_{1}$, we get
\begin{eqnarray}
\dot L=-2N^{2}\gamma\int_{0}^{t}d\tau\int d^{2}{\bf r}d^{2}{\bf r}_{1}\lbrack
{\bf r}_{1}.{\bf V}(0\rightarrow 1)\rbrack_{t} \biggl\lbrace {\bf
V}(1\rightarrow 0)P_{1}\nabla P+{\bf V}(0\rightarrow 1)P\nabla
P_{1}\biggr\rbrace_{t-\tau}.
\label{c6}
\end{eqnarray}
Adding these two quantities, we arrive at the final expression
\begin{eqnarray}
\dot L=-N^{2}\gamma\int_{0}^{t}d\tau\int d^{2}{\bf r}d^{2}{\bf r}_{1}\lbrack
{\bf r}.{\bf V}(1\rightarrow 0)+ {\bf r}_{1}.{\bf V}(0\rightarrow
1)\rbrack_{t}\nonumber\\ 
\times \biggl\lbrace {\bf V}(1\rightarrow 0)P_{1}\nabla
P+{\bf V}(0\rightarrow 1)P\nabla P_{1}\biggr\rbrace_{t-\tau}.
\label{c7}
\end{eqnarray}
Now, the term in brackets vanishes as shown by the following argument.
Consider two point vortices in a circular (or infinite) domain. Their angular
momentum is  
\begin{eqnarray}
L=\gamma (r^{2}+r_{1}^{2})
\label{c8}
\end{eqnarray}
and it is conserved. This implies:
\begin{eqnarray}
0={dL\over dt}=2\gamma \biggl ({\bf r}.{d{\bf r}\over dt}+{\bf r}_{1}.{d{\bf
r}_{1}\over dt}\biggr )=2\gamma\lbrack {\bf r}.{\bf V}(1\rightarrow 0)+ {\bf
r}_{1}.{\bf V}(0\rightarrow 1)\rbrack.
\label{c9}
\end{eqnarray}
We can also prove this result by a direct calculation. In an unbounded domain
${\bf V}(0\rightarrow 1)=-{\bf V}(1\rightarrow 0)$ and consequently
\begin{eqnarray}
{\bf r}.{\bf V}(1\rightarrow 0)+ {\bf r}_{1}.{\bf V}(0\rightarrow 1)={\bf
V}(1\rightarrow 0).({\bf r}-{\bf r}_{1})=0,
\label{c10}
\end{eqnarray}
where we have used equation (\ref{Vji}) to get the last equality. In a circular
domain, the velocity ${\bf V}(1\rightarrow 0)$ is given by equation (\ref{Vji})
{\it plus} a term ${\bf V}_{b}(1\rightarrow 0)$ which can be determined with the
method of ``images''. If $R$ denotes the domain radius, we find
\begin{eqnarray}
{\bf V}_{b}(1\rightarrow 0)={\gamma\over 2\pi}\hat {\bf z}\wedge {{R^{2}\over
r_{1}^{2}}{\bf r}_{1}-{\bf r}\over \bigl |{R^{2}\over r_{1}^{2}}{\bf r}_{1}-{\bf
r}\bigr |^{2}}.
\label{c11}
\end{eqnarray}
Therefore,
\begin{eqnarray}
{\bf r}.{\bf V}_{b}(1\rightarrow 0)+{\bf r}_{1}.{\bf V}_{b}(0\rightarrow 1)
={\gamma\over 2\pi}\Biggl \lbrace {(\hat {\bf z}\wedge {\bf r}_{1}).{\bf r}
\over \bigl |{R\over r_{1}}{\bf r}_{1}-{r_{1}\over R}{\bf r}\bigr |^{2}}+
{(\hat {\bf z}\wedge {\bf r}).{\bf r}_{1} \over \bigl |{R\over r}{\bf r}-{r\over
R}{\bf r}_{1}\bigr |^{2}}\Biggr\rbrace.
\label{c12}
\end{eqnarray}
Noting that $(\hat {\bf z}\wedge {\bf r}_{1}).{\bf r}=-(\hat {\bf z}\wedge {\bf
r}).{\bf r}_{1}$ and that
\begin{eqnarray}
\biggl |{R\over r_{1}}{\bf r}_{1}-{r_{1}\over R}{\bf r}\biggr
|^{2}=R^{2}+{r_{1}^{2}\over R^{2}}r^{2}-2{\bf r}.{\bf r}_{1}= \biggl |{R\over
r}{\bf r}-{r\over R}{\bf r}_{1}\biggr |^{2},
\label{c13}
\end{eqnarray}
we finally conclude that
\begin{eqnarray}
 {\bf r}.{\bf V}(1\rightarrow 0)+ {\bf r}_{1}.{\bf V}(0\rightarrow 1)=0.
\label{c14}
\end{eqnarray}
From this identity and from equation (\ref{c7}), it results that the kinetic
equation (\ref{new1}) conserves the angular momentum in a disk and in an
infinite domain, i.e. $\dot L=0$. 

(ii) In an infinite domain or in a channel, the linear impulse
\begin{eqnarray}
{\bf P}=\int {\bf r}\wedge \langle\omega\rangle\hat {\bf z}  d^{2}{\bf r}
\label{c15}
\end{eqnarray}
must be conserved (in a channel extending in the $x$ direction, only the
component $P_{x}$ of the linear impulse must be conserved). Taking the time
derivative of equation (\ref{c15}), substituting for (\ref{c1}) and remembering
that ${\bf P}$ is conserved by the advective term, we get
\begin{eqnarray}
\dot {\bf P}_{\perp}=N\gamma\int {\bf J} d^{2}{\bf r}.
\label{c16}
\end{eqnarray}
Substituting explicitly for the diffusion current (\ref{c2}) in equation
(\ref{c16}), permutting the dummy variables ${\bf r}$ and ${\bf r}_{1}$, and
taking the half-sum of the resulting expressions we finally obtain 
\begin{eqnarray}
\dot {\bf P}_{\perp}=-{N^{2}\gamma\over 2}\int_{0}^{t}d\tau\int d^{2}{\bf
r}d^{2}{\bf r}_{1}\lbrack {\bf V}(1\rightarrow 0)+ {\bf V}(0\rightarrow
1)\rbrack_{t}\nonumber\\ 
\times\biggl\lbrace {\bf V}(1\rightarrow 0)P_{1}\nabla
P+{\bf V}(0\rightarrow 1)P\nabla P_{1}\biggr\rbrace_{t-\tau}.
\label{c17}
\end{eqnarray}
Now, we can use the same argument as before to show that the term in brackets
vanishes. Let us consider two point vortices in a channel (or in an infinite
domain). Their linear impulse is
\begin{eqnarray}
{\bf P}_{\perp}=\gamma ({\bf r}+{\bf r}_{1})
\label{c18}
\end{eqnarray}
and it is conserved. This implies
\begin{eqnarray}
{\bf 0}={d{\bf P}_{\perp}\over dt}=\gamma \biggl ({d{\bf r}\over dt}+{d{\bf
r}_{1}\over dt}\biggr )=\gamma\lbrack {\bf V}(1\rightarrow 0)+ {\bf
V}(0\rightarrow 1)\rbrack.
\label{c19}
\end{eqnarray}
We can also prove this result by a direct calculation. Equation (\ref{c19}) is
obvious in an unbounded domain since ${\bf V}(0\rightarrow 1)=-{\bf
V}(1\rightarrow 0)$. In a channel extending in the $x$ direction, we need to
show that $\dot P_{x}=0$, i.e. $V_{y}(1\rightarrow 0)+V_{y}(0\rightarrow 1)=0$.
Now, the velocity $V_{y}(1\rightarrow 0)$ is given by equation (\ref{Vji}) {\it
plus} a term $V_{b}(1\rightarrow 0)_{y}$ which can be determined with the method
of ``images''. If $a$ denotes the width of the channel, we find that 
\begin{eqnarray}
V_{b}(1\rightarrow 0)_{y}=-{\gamma\over
2\pi}\sum_{n=-\infty}^{+\infty}\biggl\lbrace {x-x_{1}\over
(x-x_{1})^{2}+(y-y_{1}-2na)^{2}}- {x-x_{1}\over
(x-x_{1})^{2}+(y+y_{1}-2na)^{2}}\biggr\rbrace.
\label{c20}
\end{eqnarray}
Under this form, it is clear that $V_{b}(1\rightarrow 0)_{y}$ is antisymmetric
under the exchange of $1$ and $0$, so that finally
\begin{eqnarray}
V_{y}(1\rightarrow 0)+V_{y}(0\rightarrow 1)=0.
\label{c21}
\end{eqnarray}
From this identity and from equation (\ref{c17}), it results that the kinetic
equation (\ref{new1}) conserves the linear impulse in a channel or in an
infinite domain, i.e. $\dot {\bf P}={\bf 0}$. 

\section{Calculation of the memory function}
\label{sec_memory}

In this Appendix, we calculate the memory function that occurs in equation
(\ref{new1}).

\subsection{Axisymmetrical flow}
\label{sec_Gaxi}

If we assume that $P=P(r,t)$, then equation (\ref{new1}) simplifies in 
\begin{eqnarray}
{\partial P\over\partial t}=-{1\over r}{\partial\over\partial r}(r J_{r}),
\label{ww1}
\end{eqnarray}
where 
\begin{eqnarray}
J_{r}=-N\int_{0}^{t}d\tau\int d^{2}{\bf r}_{1}{V}_{r(t)}(1\rightarrow
0)_{t}\biggl\lbrace {V}_{r(t-\tau)}(1\rightarrow 0)P_{1} {\partial
P\over\partial r}-{V}_{r_{1}(t-\tau)}(1\rightarrow 0)P {\partial
P_{1}\over\partial r_{1}}\biggr\rbrace_{t-\tau}
\label{ww2}
\end{eqnarray}
and where $V_{r(t)}(1\rightarrow 0)$ is the component of the vector ${\bf
V}(1\rightarrow 0)$ in the direction of ${\bf r}(t)$. If we denote by
$(r(t),\theta(t))$ and $(r_{1}(t),\theta_{1}(t))$ the polar coordinates that
specify the position of the point vortices $0$ and $1$  at time $t$, we easily
find that
\begin{eqnarray}
V_{r(t)}(1\rightarrow 0)=-{\gamma\over 2\pi}{r_{1}\sin (\theta-\theta_{1})\over
r_{1}^{2}+r^{2}-2 r r_{1}\cos (\theta-\theta_{1})}. 
\label{ww3}
\end{eqnarray}   
We shall assume that between $t$ and $t-\tau$, the point vortices follow
circular trajectories with angular velocity $\Omega(r,t)$. In that case,
$r(t-\tau)=r$ and $\theta(t-\tau)=\theta-\Omega(r,t)\tau$. Then, we obtain
\begin{eqnarray}
V_{r(t-\tau)}(1\rightarrow 0)=-{\gamma\over 2\pi}{r_{1}\sin
(\theta-\theta_{1}-\Delta\Omega\tau)\over r_{1}^{2}+r^{2}-2 r r_{1}\cos
(\theta-\theta_{1}-\Delta\Omega\tau)} 
\label{ww5}
\end{eqnarray}  
with  
\begin{eqnarray}
\Delta\Omega=\Omega(r,t)-\Omega(r_{1},t).
\label{ww4}
\end{eqnarray} 
We find similarly that $V_{r_{1}(t-\tau)}(1\rightarrow 0)={r\over
r_{1}}V_{r(t-\tau)}(1\rightarrow 0)$. Our previous assumptions also imply that
$P(r(t-\tau),t-\tau)\simeq P(r,t)$ between $t$ and $t-\tau$. In words, this
means that the correlation time is smaller than the time scale on which the
average vorticity changes appreciably. We do not assume that it is {\it much}
smaller, so this approximation is not over restrictive. In that case, the diffusion
current becomes 
\begin{eqnarray}
J_{r}=-N\int_{0}^{+\infty}d\tau\int_{0}^{2\pi}d\theta_{1}\int_{0}^{+\infty}r
r_{1}dr_{1} {V}_{r(t)}(1\rightarrow 0) {V}_{r(t-\tau)}(1\rightarrow 0)
\biggl\lbrack {1\over r}P_{1}{\partial P\over\partial r}- {1\over
r_{1}}P{\partial P_{1}\over\partial r_{1}}\biggr\rbrack, 
\label{ww6}
\end{eqnarray}
where the time integral has been extended  to $+\infty$. We now need to evaluate
the memory function
\begin{eqnarray}
M=\int_{0}^{+\infty}d\tau\int_{0}^{2\pi}d\theta_{1} {V}_{r(t)}(1\rightarrow 0)
{V}_{r(t-\tau)}(1\rightarrow 0).
\label{ww7}
\end{eqnarray}
Introducing the notations $\phi=\theta_{1}-\theta$ and 
\begin{eqnarray}
\lambda={2 r r_{1}\over r_{1}^{2}+r^{2}}<1,
\label{ww8}
\end{eqnarray}
we have explicitly:
\begin{eqnarray}
M=\biggl ({\gamma\lambda\over 4\pi r}\biggr
)^{2}\int_{0}^{+\infty}d\tau\int_{0}^{2\pi}d\phi {\sin\phi\over
1-\lambda\cos\phi}\  {\sin(\phi+\Delta\Omega\tau)\over 1-\lambda\cos
(\phi+\Delta\Omega\tau)}.
\label{ww9}
\end{eqnarray}
This can also be written
\begin{eqnarray}
M=\biggl ({\gamma\over 4\pi r}\biggr
)^{2}\int_{0}^{+\infty}d\tau\int_{0}^{2\pi}d\phi V'(\phi)
V'(\phi+\Delta\Omega\tau), 
\label{ww10}
\end{eqnarray}
where 
\begin{eqnarray}
V(\phi)=\ln (1-\lambda\cos\phi).
\label{ww11}
\end{eqnarray}
We now write the function $V(\phi)$ in the form of a Fourier series
\begin{eqnarray}
V(\phi)=\sum_{n=-\infty}^{+\infty}a_{n}e^{i n\phi}\qquad {\rm with}\qquad
a_{n}={1\over 2\pi}\int_{-\pi}^{\pi} V(\phi)e^{-i n \phi}d\phi.
\label{ww12}
\end{eqnarray} 
The memory function becomes
\begin{eqnarray}
M=-{1\over 2}\biggl ({\gamma\over 4\pi r}\biggr
)^{2}\int_{-\infty}^{+\infty}d\tau\int_{0}^{2\pi}d\phi
\sum_{n,m=-\infty}^{+\infty}n m a_{n}a_{m}e^{i(n+m)\phi}e^{i m \Delta
\Omega\tau}.
\label{ww14}
\end{eqnarray}
Carrying out the integrations on $\phi$ and $\tau$ using the integral
representation of the delta function
\begin{eqnarray}
\delta (x)={1\over 2\pi}\int_{-\infty}^{+\infty} e^{-i\rho x}d\rho,
\label{wwr14}
\end{eqnarray}
we are left with
\begin{eqnarray}
M=-{\gamma^{2}\over 8 r^{2}} \sum_{n,m=-\infty}^{+\infty}n m
a_{n}a_{m}\delta_{n,-m} \delta(m\Delta\Omega)={\gamma^{2}\over 8 r^{2}}
\delta(\Delta\Omega)  \sum_{n=-\infty}^{+\infty} |n| a_{n}^{2} . 
\label{ww15}
\end{eqnarray}
It remains for us to evaluate the series that appears in the last expression of
the memory function. Using the identities:
\begin{eqnarray}
\int_{0}^{\pi}\ln (1-\lambda\cos\phi)\cos(n\phi)d\phi=-{\pi\over n}\biggl
({1\over\lambda}-\sqrt{{1\over\lambda^{2}}-1}\biggr )^{n} \qquad (n>0) 
\label{ww18}
\end{eqnarray}
\begin{eqnarray}
\int_{0}^{\pi}\ln (1-\lambda\cos\phi)d\phi=\pi\ln\biggl ({1\over
2}+{\sqrt{1-\lambda^{2}}\over 2}\biggr )
\label{ww19}
\end{eqnarray}
and the definition (\ref{ww8}) of $\lambda$, we find that $a_{0}<\infty$ and,
for $n>0$:
\begin{eqnarray}
a_{n}=-{1\over n}\biggl ({(r_{1}^{2}+r^{2})-|r_{1}^{2}-r^{2}|\over 2 r
r_{1}}\biggr )^{n}=-{1\over n}\biggl ({r_{<}\over r_{>}}\biggr )^{n},
\label{ww20}
\end{eqnarray}
where $r_{>}$ (resp. $r_{<}$) is the biggest (resp. smallest) of $r$ and
$r_{1}$. Therefore, the value of the series is
\begin{eqnarray}
\sum_{n=-\infty}^{+\infty} |n| a_{n}^{2}= 2\sum_{n=1}^{+\infty} n a_{n}^{2}=
2\sum_{n=1}^{+\infty} {1\over n}\biggl ({r_{<}\over r_{>}}\biggr )^{2n}=-2
\ln\biggl \lbrack 1-\biggl ({r_{<}\over r_{>}}\biggr )^{2}\biggr \rbrack .
\label{ww21}
\end{eqnarray}
The memory function takes the form
\begin{eqnarray}
M=-{\gamma^{2}\over 4 r^{2}} \delta(\Delta\Omega)\ln\biggl \lbrack 1-\biggl
({r_{<}\over r_{>}}\biggr )^{2}\biggr \rbrack    
\label{ww22}
\end{eqnarray}
and the diffusion current in the axisymmetrical case can be written
\begin{eqnarray}
J_{r}={N\gamma^{2}\over 4 r}\int_{0}^{+\infty}r_{1}dr_{1}
\delta(\Omega-\Omega_{1})\ln\biggl \lbrack 1-\biggl ({r_{<}\over r_{>}}\biggr
)^{2}\biggr \rbrack\biggl\lbrace {1\over r}P_{1}{\partial P\over\partial r}-
{1\over r_{1}}P{\partial P_{1}\over\partial r_{1}} \biggr\rbrace.   
\label{ww23}
\end{eqnarray}
This leads to the kinetic equation (\ref{day1}).

\subsection{Unidirectional flow}
\label{sec_Guni}

If we assume that $P=P(y,t)$, then equation (\ref{new1}) simplifies in
\begin{eqnarray}
{\partial P\over\partial t}=-{\partial J_{y}\over\partial y}
\label{ur1}
\end{eqnarray}
with
\begin{eqnarray}
J_{y}=-N\int_{0}^{t}d\tau\int dx_{1}dy_{1}V^{y}(1\rightarrow
0)_{t}V^{y}(1\rightarrow 0)_{t-\tau}\biggl\lbrace P_{1}{\partial P\over\partial
y}- P  {\partial P_{1}\over\partial y_{1}}\biggr \rbrace_{t-\tau}.
\label{u1}
\end{eqnarray}
Assuming that between $t$ and $t-\tau$ the vortices follow linear trajectories
with velocity $\langle {\bf V}\rangle=\langle V\rangle (y,t) {\bf e}_{x}$, we
have $y(t-\tau)=y$ and $x(t-\tau)=x-\langle V\rangle (y,t) \tau$. Therefore, the
function $V^{y}(1\rightarrow 0)$ at times $t$ and $t-\tau$ takes explicitly the
form
\begin{eqnarray}
V^{y}(1\rightarrow 0)_{t}=-{\gamma\over 2\pi}{x_{1}-x\over
(x_{1}-x)^{2}+(y_{1}-y)^{2}}
\label{u3}
\end{eqnarray}
and 
\begin{eqnarray}
V^{y}(1\rightarrow 0)_{t-\tau}=-{\gamma\over 2\pi}{x_{1}-x-\Delta V \tau \over
(x_{1}-x-\Delta V\tau)^{2}+(y_{1}-y)^{2}},
\label{u4}
\end{eqnarray}
where we have introduced the notation
\begin{eqnarray}
\Delta V=\langle V\rangle (y_{1},t)-\langle V\rangle (y,t).
\label{u5}
\end{eqnarray}
We also assume that the correlation time is
smaller than the time scale over which the vorticity changes appreciably. Then,
$P(y(t-\tau),t-\tau)\simeq P(y,t)$ and the diffusion current becomes
\begin{eqnarray}
J_{y}=-N\int_{0}^{t}d\tau\int dx_{1}dy_{1}V^{y}(1\rightarrow
0)_{t}V^{y}(1\rightarrow 0)_{t-\tau}\biggl\lbrace P_{1}{\partial P\over\partial
y}- P  {\partial P_{1}\over\partial y_{1}}\biggr \rbrace.
\label{ue1}
\end{eqnarray}
We now need to calculate the memory function 
\begin{eqnarray}
M=\int_{0}^{+\infty}d\tau\int_{-\infty}^{+\infty} dx_{1} V^{y}(1\rightarrow
0)_{t}V^{y}(1\rightarrow 0)_{t-\tau}.
\label{u2}
\end{eqnarray}
Using equations (\ref{u3}) and (\ref{u4}), we have explicitly
\begin{eqnarray}
M={\gamma^{2}\over 4\pi^{2}}\int_{0}^{+\infty}d\tau\int_{-\infty}^{+\infty} dX
{X\over X^{2}+Y^{2}}{X-\Delta V \tau \over (X-\Delta V\tau)^{2}+Y^{2}},
\label{u6}
\end{eqnarray}
where we have set $X=x_{1}-x$, $Y=y_{1}-y$. Equation (\ref{u6}) can also be
written
\begin{eqnarray}
M={\gamma^{2}\over 4\pi^{2}}\int_{0}^{+\infty}d\tau\int_{-\infty}^{+\infty} dX
{\partial W\over\partial X}(X,Y){\partial W\over\partial X}(X-\Delta V\tau,Y)
\label{u7}
\end{eqnarray}
with
\begin{eqnarray}
W(X,Y)=\ln\sqrt{X^{2}+Y^{2}}=\ln \xi
\label{u8}
\end{eqnarray}
We shall now write the function $W({\mb\xi})$ in the form of a Fourier integral
\begin{eqnarray}
W({\mb\xi})={1\over (2\pi)^{2}}\int \hat W({\bf k})e^{-i{\bf
k}{\mb\xi}}d^{2}{\bf k}\qquad {\rm with }\qquad \hat W({\bf k})=\int
W({\mb\xi})e^{i{\bf k}{\mb\xi}}d^{2}{\mb \xi}.
\label{u9}
\end{eqnarray}
Then, the foregoing expression for the memory function becomes
\begin{eqnarray}
M=-{\gamma^{2}\over (2\pi)^{6}}\int_{0}^{+\infty}d\tau\int_{-\infty}^{+\infty}
dX \int d^{2}{\bf k}d^{2}{\bf k}' \hat W({\bf k})\hat W({\bf k}')k_{x}k_{x}'
e^{-i({\bf k}+{\bf k}'){\mb\xi}}e^{i k_{x}'\Delta V\tau}.
\label{u10}
\end{eqnarray}
Carrying out the integrations over $\tau$ and $X$, we get
\begin{eqnarray}
M=-{\gamma^{2}\over 32\pi^{4}} \int d^{2}{\bf k}d^{2}{\bf k}' \hat W({\bf
k})\hat W({\bf k}')k_{x}k_{x}' \delta (k_{x}+k_{x}')
e^{-i({k}_{y}+{k}'_{y})Y}\delta(k_{x}'\Delta V)
\label{u11}
\end{eqnarray}
and, consequently,
\begin{eqnarray}
M={\gamma^{2}\over 32\pi^{4}}\delta(\Delta V) \int_{-\infty}^{+\infty} dk_{x}
dk_{y} dk'_{y}\hat W(k_{x},k_{y})\hat W(-k_{x},k_{y}')|k_{x}|
e^{-i({k}_{y}+{k}'_{y})Y}.
\label{u12}
\end{eqnarray}
Now, the Fourier transform of $W$ can be written explicitly
\begin{eqnarray}
\hat W({\bf k})=2\pi \int_{0}^{+\infty} W(\xi)J_{0}(k\xi)\xi d\xi,
\label{u13}
\end{eqnarray}
where use has been made of the well-known identity
\begin{eqnarray}
\int_{0}^{2\pi} \cos(z\cos\theta)d\theta=2\pi J_{0}(z),
\label{u14}
\end{eqnarray}
where $J_{0}$ is Bessel function of order zero. It is immediate to see that the
Fourier transform of $W$ as defined by (\ref{u8}) does not exists. Indeed, the
integral (\ref{u13}) diverges when $\xi\rightarrow +\infty$, i.e. at large
separations. However, in physical situations the domain never extends to
infinity so that, in practice, the integral remains finite. A convenient way to
introduce a cut-off at large separations is to make the substitution
\begin{eqnarray}
W({\mb\xi})=\ln \xi \qquad \rightarrow \qquad W({\mb\xi})=-K_{0}(\xi/L),
\label{u15}
\end{eqnarray} 
where $K_{0}$ is the modified Bessel function of order zero and  $L$ is a length
scale of the order of the system size. For small separations
$K_{0}(z)\sim -\ln z$ and for large separations $K_{0}(z)\sim \sqrt{\pi\over
2z}e^{-z}$. This modification amounts to replacing the Poisson equation
(\ref{poisson}) by an equation of the form
\begin{eqnarray}
-\Delta\psi+{1\over L}\psi=\omega
\label{u16}
\end{eqnarray} 
Equation (\ref{u16}) is precisely what is obtained in geophysics in the
``quasigeostrophic approximation''. The deformation of the fluid surface
introduces a shielding of the interaction between vortices on a length $\sim L$,
called the Rossby radius. Obviously, the Rossby radius plays the same role as
the Debye length in plasma physics. 

With this prescription we find that
\begin{eqnarray}
\hat W({\bf k})=-{2\pi \over k^{2}+k_{L}^{2}}
\label{u17}
\end{eqnarray}
where $k_{L}=1/L$. Substituting in equation (\ref{u12}), we get
\begin{eqnarray}
M={\gamma^{2}\over 2\pi^{2}}\delta(\Delta V) \int_{-\infty}^{+\infty}
dk_{x}|k_{x}|\biggl (\int_{0}^{+\infty} { \cos(k_{y}Y)\over
k_{L}^{2}+k_{x}^{2}+k_{y}^{2}}dk_{y}\biggr )^{2}.
\label{u18}
\end{eqnarray}
The integral on $k_{y}$ can be carried out easily, leaving the result
\begin{eqnarray}
M={\gamma^{2}\over 4}\delta(\Delta V) \int_{0}^{+\infty} dk_{x}{k_{x}\over
k_{L}^{2}+k_{x}^{2}}e^{-2|Y|\sqrt{ k_{L}^{2}+k_{x}^{2}}}.
\label{u19}
\end{eqnarray}
Now, setting $t^{2}=4Y^{2}(k_{x}^{2}+k_{L}^{2})$, we finally obtain 
\begin{eqnarray}
M={\gamma^{2}\over 4}\delta(\Delta V)E_{1}\biggl ({2|Y|\over L}\biggr ), 
\label{u20}
\end{eqnarray}
where
\begin{eqnarray}
E_{1}(x)=\int_{x}^{+\infty}{e^{-t}\over t}dt
\label{u21}
\end{eqnarray}
is the exponential integral. For $x\rightarrow 0$, we have $E_{1}(x)=-C-\ln x$
where $C=0.57721...$ is Euler's constant. In conclusion, the diffusion current
in the unidirectional case takes the form
\begin{eqnarray}
J_{y}=-{N\gamma^{2}\over 4}\int_{-\infty}^{+\infty} dy_{1}\delta (
V_{1}-V)E_{1}\biggl ({2|y_{1}-y|\over L}\biggr )\biggl\lbrace P_{1}{\partial
P\over\partial y}- P{\partial P_{1}\over\partial y_{1}}\biggr\rbrace
\label{u22}
\end{eqnarray}
and it leads to the kinetic equation (\ref{day5}).

\end{document}